\documentclass[12pt,a4paper]{article}
\usepackage{amsfonts}
\usepackage{mathrsfs}  
\usepackage{epsfig}
\usepackage{axodraw2}
\usepackage{graphicx}
\usepackage{url}

\begin{document}

\title{Free energy of the gas of spin-1/2 fermions beyond the second order and the Stoner phase transition}
\author{\em Oskar Grocholski$~\!^1$ and Piotr H.
  Chankowski$~\!^2$\footnote{Emails:
    oskar.grocholski@cea.fr, chank@fuw.edu.pl}\\
  $^1$IRFU, CEA, Universit\'e Paris-Saclay, F-91191 Gif-sur-Yvette, France\\
$^2$Faculty of Physics, University of Warsaw,\\
Pasteura 5, 02-093 Warsaw, Poland
}
\maketitle
\abstract{Applying the previously developed systematic thermal (imaginary time)
  perturbative expansion to the relevant effective field theory we compute the
  free energy $F$ of the diluted gas of (nonrelativistic) spin $1/2$ fermions
  interacting through a spin-independent repulsive two-body potential as a
  function of the numbers $N_+$ and $N_-$ of spin up and spin down fermions
  (i.e. as a function of the system's polarization) and the temperature $T$.
  We give the complete order $(k_{\rm F}a_0)^3$ ($k_{\rm F}$ is the Fermi
  wave vector and $a_0$ is the $s$-wave scattering length characterizing
  the interaction potential) contribution to $F$.
  We also extend the computation beyond a fixed order by resumming to all
  orders in the parameter $k_{\rm F}a_0$ the contributions to $F$ of two
  infinite sets of Feynman
  diagrams: the so-called particle-particle rings and the particle-hole rings.
  We find that including the second one of these two contributions
  has a dramatic consequence for   the transition of the system from the
  paramagnetic to the ferromagnetic phase (the so called Stoner phase
  transition): in this approximation the phase transition simply disappears.
 This result does not contradict the expectation that a transition
  to the magnetically ordered state should occur in truly repulsive
  systems. The $p$-wave and higher scattering lengths, as well as other
  parameters chacterizing the interaction potential, are in such systems 
  generally of the same order of magnitude as $a_0$ and contributions
  depending on them should be, therefore, also included in $F$. Our
  results may, however, have implications for the search of the
  itinerant ferromagnetism of cold atomic gases in which large $a_0$, much
  larger than all other parameters, is artificially created by exploiting
  the physics of the Feshbach resonance.} 
\vskip0.1cm


\newpage

\noindent{\large\bf 1. Introduction}
\vskip0.3cm

\noindent In recent years some progress has been achieved in the computation
of equilibrium zero temperature properties of the gas of $N$ (nonrelativistic)
fermions interacting with each other through a spin independent repulsive
potential $V_{\rm pot}(|{\mathbf x}_i-{\mathbf x}_j|)$. It was mainly related to
the application to this classic
\cite{Lenz,Stoner,HuangYang57,Kesio,Pathria,FetWal}
many-body quantum mechanics and statistical physics problem of general methods (see
e.g. \cite{KolczastyiSka}) of the effective field theory. In this approach,
initiated in the seminal paper \cite{HamFur00} (see also \cite{HamFur02}), the
original spatially nonlocal potential $V_{\rm pot}$ assumed to be characterized
by a length scale $R$ is replaced by the \textit{a priori} infinite series of local
interactions (written here using the standard second quantization formalism -
see e.g. Ref. \cite{FetWal} - with $\hat\psi_\pm$ being the field operators of spin
up and down fermions)
\begin{eqnarray}
  \hat V_{\rm int}=C_0\!\int\!d^3{\mathbf x}~\!
  (\hat\psi_+^\dagger\hat\psi_+)(\hat\psi_-^\dagger\hat\psi_-)
  +\hat V_{\rm int}^{(C_2)}+\hat V_{\rm int}^{(C_2^\prime)}+\dots,
  \label{eqn:Vint}
\end{eqnarray}
of decreasing length dimension. The coefficients (couplings) $C_0$, $C_2$,
$C_2^\prime,\dots$ of the interaction (\ref{eqn:Vint}) can be then directly
determined in terms of the quantities - the scattering lengths $a_0, a_1,\dots$
and the effective radii $r_0,\dots$ - parametrizing the general expansion (in
powers of the relative momentum) of the amplitude of the elastic scattering of
two particles. Trading the (bare) couplings of (\ref{eqn:Vint}) for these
measurable quantities which characterize the underlying potential $V_{\rm pot}$
of the binary interactions also has the effect of removing UV
infinities engendered by the locality of (\ref{eqn:Vint}). The simplifications
brought in by this approach allowed first to easily reproduce \cite{HamFur00}
those terms of the perturbative expansion of the (ground-state) energy density
$E/V$ of the system of spin $s$ fermions with equal densities of different spin
projections which in the past were obtained by more traditional (and requiring
considerably more work) methods of many-body quantum mechanics
 (such as the ones based on Green's functions or on the Galitskii approach)
\cite{FetWal} and to extend \cite{WeDrSch} this computation
to the fourth order in the systematic, organized by the power counting rules
\cite{HamFur00},
expansion in powers (in higher orders modified also by logarithms) of the
dimensionless product $k_{\rm F}R$ of the system's overall Fermi wave vector
\begin{eqnarray}
  k_{\rm F}=(6\pi^2n/g_s)^{1/3}~\!,\label{eqn:kFDef}
\end{eqnarray}
(here $n=N/V$ is the overall density of the gas of spin $s$ fermions and
$g_s=2s+1$) and the characteristic length $R$. The same approach allowed also
to compute up to order $(k_{\rm F}a_0)^2$ the ground state energy of the system of spin $1/2$ fermions for
different densities of fermions with the
up and down spin projections \cite{CHWO1} recovering the old result of Kanno
\cite{KANNO} (obtained analytically for the hard core interaction potential)
and to easily extend it \cite{PECABO1} to fermions having spin $s$ greater
than 1/2 (and, thus, more possible spin projections).
Finally, in Refs. \cite{CHWO3,CHWO4} and slightly later in \cite{PECABO2} this
  computation has been extended up to
the third order in the systematic expansion in powers of $k_{\rm F}R$.

These technical achievements allowed to investigate more quantitatively
the phase transition, also
called the Stoner transition, to the ordered state in which the densities
$N_+/V$ and $N_-/V$ of fermions of opposite spin projections in the case of
$s=1/2$ are not equal and the system exhibits a nonzero polarization
$P\equiv(N_+-N_-)/N$. According to the standard qualitative argument
based on the positivity of the repulsive interaction energy and the Pauli 
exclusion principle, for sufficiently strong repulsion and/or sufficiently
high overall density, such an ordered state should be energetically (at zero
temperature) favored
over the state with equal densities. In this regime the system should,
therefore, exhibit the property called itinerant ferromagnetism. It has been
found that the clear first order character of this transition 
predicted by the second order of the perturbative expansion (in
agreement with the general quantitative arguments given long ago in Ref. 
\cite{BeKiVoj}) - at sufficiently large values of the ``gas parameter''
$k_{\rm F}a_0$ two symmetric minima of $E(P)$ are formed away from $P=0$
separated from the one at $P=0$ by a finite barrier and at
$(k_{\rm F}a_0)_{\rm cr}=1.054$ they become the global minima
($|P_{\rm cr}|=0.58$) - gets significantly weakened ($|P_{\rm cr}|$ shifts
significantly toward $P=0$ and the barrier becomes much lower) if the complete
order $(k_{\rm F}a_0)^3$ contribution is taken into account \cite{CHWO3,CHWO4}.
This seemed to be in line with the results of the work \cite{He1} in which a
certain class of contributions to $E/V$ (arising from the so-called
particle-particle ring diagrams) has been resummed to all orders in
$k_{\rm F}a_0$ finding that in this approximation the
transition is continuous - the new minima start to continuously move
away from $P=0$ as the gas parameter crosses some critical value
$(k_{\rm F}a_0)_{\rm cr}\approx0.8$ ($0.858$ if only 1 hole-hole $N-1$
particle-particle parts of the $N$-th order particle-particle ring
diagrams and $0.79$ if the complete $N$-th order particle-particle ring
diagrams are resummed; see also Ref. \cite{He2} for a refinement of this approach).
It should be added that the predictions of Ref. \cite{He1}, which qualitatively
seem to be supported by the results of a different approach \cite{HEIS},
agree at first sight quite well with the results
obtained with the help of the Monte Carlo simulations \cite{QMC10}.

However if all scattering lengths $a_\ell$ and effective radii $r_\ell$ are
of the same order of magnitude $\sim\!R$, the complete (according to the power
counting rules \cite{HamFur00} which apply, strictly speaking, only to such a
case) order $(k_{\rm F}R)^3$ contribution to the energy density depends
also on $a_1$ and $r_0$ and it has been shown in Ref. \cite{PECABO2} that in this
case the character of the phase transition (at zero temperature) predicted
by this approximation depends on the relative magnitudes and signs of the
parameters $a_1$, $r_1$ and $a_0>0$ (that is, on the more detailed
characteristics of the underlying potential $V_{\rm pot}$). The computation
 (such as the one done in Ref. \cite{He1}) in which to all orders resummed are
(some) contributions depending only on powers of $k_{\rm F}a_0$ correspond
rather to the situation encountered in physics of dilute atomic gases in which
the $s$-wave scattering length $a_0$ (it can be of either sign) is made
positive and very large compared to the remaining parameters
($|a_0|\gg R\sim |a_1|,|r_1|,\dots$) by exploiting properties of the Feshbach
resonances (see e.g. Ref. \cite{ChiGriJuTie}). In this case however the underlying
interaction of fermions (atoms) is attractive and the scattering length $a_0$
is positive in the regime in which bound states composed of two fermions
of opposite spins can form. The true ground state of the system is then very
different from the one of nointeracting atoms (they may not be adiabatically
connected to one another in the thermodynamic limit implicit in the field
theory approach) which is used in perturbative computations performed within
the effective field theory approach.
In view of that the energy density $E/V$ obtained perturbatively by
    expanding around the ground-state of the system of noninteracting
    fermions most probably is the energy density of the metastable (from
    the thermodynamic point of view) state in which a real gas of cold
    fermionic atoms finds itself just after switching the external magnetic
    field so that the $s$-wave scattering length $a_0$ becomes positive and
    large. However, although the dependence on the polarization
    of this energy density indicates that the transition to the
    ferromegnetic state (which in principle could occur in a metastable state
    \cite{Pippard}) should occur, a too short lifetime of the gas metastable state
    may prevent its observation in real systems.
 Therefore the failure of the
  attempts to observe such a transition \cite{ItFMObs,ItFMNotObsT,ItFMNotObsE}
  in experiments is usually attributed to the too rapid formation (at very
  low temperatures at which these experiments were carried out) of atomic
  dimers (bosons). It should however be remarked
that in the situation in which the underlying interaction is attractive
(despite giving rise to a positive $s$-wave scattering length)
the mentioned qualitative argument for the occurrence of the transition
no longer applies and the expectation that the transition should occur
is mainly based on the textbook mean field correction to the energy
density \cite{Kesio,Pathria}
(equivalent to the first order correction of the perturbative expansion)
which depends only linearly on the $s$-wave scattering length $a_0$.

Computation of the ground-state energy density $E/V$ as a function of
the densities of fermions with different spin projections allows only to
investigate the equilibrium properties of the system of interacting
fermions at zero (or a very low) temperature. It is however of interest to
determine its behavior also at nonzero temperatures. (In the context of
the physics of atomic gases it  is physically clear that at higher temperatures
 the mentioned formation of atomic dimers
should be less important.) This requires computing
one of the thermodynamic potentials of the system of interacting fermions.
So far such a computation of the free energy $F$ has been
done \cite{DUMacDO} only up to the second order, i.e. up to terms of order
$(k_{\rm F}a_0)^2$, using the old-fashioned thermal perturbation theory
(see e.g. Ref. \cite{LL}, paragraph 32) based on the ordinary second  order
Rayleigh -  Schr\"odinger perturbative expression for energy levels
entering the statistical sum.  In Ref. \cite{CHGR} we have applied to the
  effective field theory (\ref{eqn:Vint}) the standard thermal perturbative
  expansion (exploiting the imaginary time formalism - see e.g. Ref. \cite{FetWal})
  which is
  routinely used to compute thermodynamic characteristics of much more complex
  systems like e.g. the quark-gluon plasma. Using it we
recovered this second order expression for $F$ and reproduced the thermal
characteristics of the Stoner phase transition it predicts (however, pointing out
 problems - not discussed in \cite{DUMacDO} - with accurate numerical
determination of the critical values
of the polarization), but have encountered
a technical problem which prevented us from immediately extending the computation
to higher orders. Here we show how this problem can be resolved and
applying the developed systematic thermal expansion to the first
term of the effective field theory interaction (\ref{eqn:Vint}) we
derive the formulas allowing to compute numerically
the complete order $(k_{\rm  F}a_0)^3$ corrections to
the free energy $F(T,V,N_+,N_-)$. With additional work, including the 
contributions of the next two terms of (\ref{eqn:Vint}) it would 
be possible to compute the complete order $(k_{\rm  F}R)^3$ correction
to the free energy.

In this paper, 
   instead of completing the order $(k_{\rm F}R)^3$ correction 
   to the free energy (which would allow to investigate in more detail
   the thermal profile of the Stoner phase transition to the ordered
   state induced by truly repulsive spin-independent two-body potetials
   $V_{\rm pot}$ which necessarily give rise to the parameters $a_1$ and $r_0$
   of comparable magnitude to that of $a_0$), we profit
 from the possibility provided by the
simpler structure of the terms of the expansion generated by the imaginary
time formalism and resum to all orders in $k_{\rm  F}a_0$ not only the
contributions to the temperature-dependent free energy $F$ of the
particle-particle ring diagrams (done for zero temperature in Refs.
\cite{He1,He2,He3}) but also of the particle-hole ring diagrams.  The
thermodynamic potential obtained in this way is most probably the free energy
of the system of fermionic atoms in a metastable state on the BEC side of the
Feshbach resonance and can therefore serve to study its properties.

Including in the free energy $F(T,V,N,P)$ only the resummed contribution
of the particle-particle ring diagrams we recover for $T=0$ the results of the
works \cite{He1,He3} and can study how they are modified at nonzero
temperatures
not exceeding the Fermi temperature (we expect that the results obtained within
the effective field theory should be valid for temperatures in this range).
However we find that the inclusion  of the contribution of the resummed
particle-hole diagrams changes the situation drastically:
the phase transition to the ordered state simply disappears (the minimum of
$F$ is at $P=0$ for all values of the parameter $k_{\rm F}a_0$ and all
temperatures). This is a somewhat surprising result and we comment on its
possible meaning
 for the problem of itinerant ferromagnetism in system of cold gases
in the Conclusions.

\vskip0.5cm

\noindent{\large\bf 2. The formalism}
\vskip0.3cm

\noindent The natural statistical formalism in which to compute equilibrium
properties of the gas of fermions the interaction of which preserves their spins
and therefore the numbers $N_\pm$ of spin up and spin down particles, is the
grand canonical ensemble with two independent chemical potentials $\mu_\pm$.
The relevant statistical operator is then
\begin{eqnarray}
  \hat\rho={1\over\Xi_{\rm stat}}~\!e^{-\beta\hat K}~\!,
\end{eqnarray}
where $\beta\equiv1/k_{\rm B}T$, with $T$ the temperature and $k_{\rm B}$
the Boltzmann constant, and 
\begin{eqnarray}
  \hat K=\hat H_0-\mu_+\hat N_+-\mu_-\hat N_-+\hat V_{\rm int}\equiv
  \hat K_0+\hat V_{\rm int}~\!.\label{eqn:KandK0}
\end{eqnarray}
The associated partition function
$\Xi_{\rm stat}(T,V,\mu_+,\mu_-)={\rm Tr}(e^{-\beta\hat K})$ gives the
thermodynamical potential $\Omega(T,V,\mu_+,\mu_-)=-Vp(T,\mu_+,\mu_-)
=-k_{\rm B}T\ln\Xi_{\rm stat}(T,V,\mu_+,\mu_-)$. In the second quantization
formalism \cite{FetWal} the operator $\hat K_0$ of the considered system
of fermions has the form\footnote{To simplify the formulas the symbol
  $\int_{\mathbf p}$ stands for the integral with respect to the measure
  $d^3{\mathbf p}/(2\pi)^3$.}
\begin{eqnarray}
  \hat K_0=\sum_{\sigma=\pm}\!\int_{\mathbf p}(\varepsilon_{\mathbf p}-\mu_\sigma)
  a^\dagger_\sigma({\mathbf p})a_\sigma({\mathbf p})~\!,
\end{eqnarray}
with $\varepsilon_{\mathbf p}=\hbar^2{\mathbf p}^2/2m_f$. The standard systematic
thermodynamical perturbative expansion \cite{FetWal,CHGR} gives the potential
$\Omega$ in the form of the series in powers of the interaction
$\hat V_{\rm int}$
\begin{eqnarray}
\Omega=\Omega^{(0)}-{1\over\beta}\sum_{N=1}^\infty{(-1)^N\over N!}
\!\int_0^\beta\!d\tau_N\dots\!\int_0^\beta\!d\tau_1~\!{\rm Tr}\!\left(
\hat\rho^{(0)}{\rm T}_\tau[\hat V_{\rm int}^I(\tau_N)\dots\hat V_{\rm int}^I(\tau_1)]
\right)^{\rm con}.\label{eqn:OmegaPertExpansion}
\end{eqnarray}
Here $\hat V_{\rm int}^I(\tau)=e^{\tau\hat K_0}\hat V_{\rm int}e^{-\tau\hat K_0}$ is the
interaction operator in the (imaginary time) interaction picture, ${\rm T}_\tau$
is the chronological ordering and $\hat\rho^{(0)}$ is the statistical operator
of the noninteracting system. The superscript ``con'' means that only connected
contributions (Feynman diagrams) should be taken into account. The first term
in (\ref{eqn:OmegaPertExpansion}) is the textbook \cite{Kesio,Pathria} grand
thermodynamical potential of the system of nointeracting fermions
\begin{eqnarray}
  \Omega^{(0)}(T,V,\mu_+,\mu_-)=-{V\over\beta}\sum_{\sigma=\pm}\int_{\mathbf p}
  \!\ln\!\left(1+e^{-\beta(\varepsilon_{\mathbf p}-\mu_\sigma)}\right).
  \label{eqn:Omega0Textbook}
\end{eqnarray}
Owing to the thermal analog of the Wick formula (see e.g. Ref. \cite{FetWal})
computation of the successive terms $\Omega^{(N)}$ with $N\geq1$
of the expansion (\ref{eqn:OmegaPertExpansion})
reduces to drawing all possible connected Feynman diagrams with $N$
interaction vertices arising from $\hat V_{\rm int}$ joined by oriented lines
and integrating over positions ${\mathbf x}$ and ``times'' $\tau$ labeling
these vertices the corresponding products of the free thermal propagators
\begin{eqnarray}
  -{\cal G}^{(0)}_{\sigma_2\sigma_1}(\tau_2-\tau_1,{\mathbf x}_2-{\mathbf x}_1)
  ={1\over\beta}\sum_{n\in{\mathbb Z}}\int_{\mathbf p}e^{-i\omega_n^{\rm F}(\tau_2-\tau_1)}
  ~\!e^{i{\mathbf p}\cdot({\mathbf x}_2-{\mathbf x}_1)}\left(
  -\tilde{\cal G}^{(0)}_{\sigma_2\sigma_1}(\omega_n^{\rm F},{\mathbf p})\right),
\end{eqnarray}
the Fourier transform $-\tilde{\cal G}^{(0)}_{\sigma_2\sigma_1}$ of which have
the form \cite{FetWal}
\begin{eqnarray}
  -\tilde{\cal G}^{(0)}_{\sigma_2\sigma_1}(\omega_n^{\rm F},{\mathbf p})
  ={-\delta_{\sigma_2\sigma_1}\over
    i\omega_n^{\rm F}-(\varepsilon_{\mathbf p}-\mu_\sigma)}~\!,
\end{eqnarray}
associated with (oriented) lines connecting vertices of each diagram.
The resulting ``momentum'' space Feynman rules are almost identical with the
ordinary ones except that integrations over frequencies (energies) are
replaced by  summations over the (fermionic) Matsubara frequencies
$\omega_n^{\rm F}=(\pi/\beta)(2n+1)$, $n\in{\mathbb Z}$.

Applying this formalism with the interaction operator $\hat V_{\rm int}$ given
by the first term of (\ref{eqn:Vint}) one finds that the order $C_0$ term
of the expansion (\ref{eqn:OmegaPertExpansion}) is simply given by
\begin{eqnarray}
  \Omega^{(1)}=C_0V{\cal G}_{++}(0,{\mathbf 0})~\!{\cal G}_{--}(0,{\mathbf 0})~\!,
  \label{eqn:Omega(1)}
\end{eqnarray}
with
\begin{eqnarray}
  {\cal G}_{\pm\pm}(0,{\mathbf 0})=\int_{\mathbf p}
  \left[1+e^{\beta(\varepsilon_{\mathbf p}-\mu_\pm)}\right]^{-1}~\!.\label{eqn:G(0,0)}
\end{eqnarray}
Higher-order contributions to the potential $\Omega$ can also be systematically
computed. If $\hat V_{\rm int}$ in (\ref{eqn:KandK0}) were the true, spatially
nonlocal, two-body interaction [corresponding to a two-body potential
$V_{\rm pot}(|{\mathbf x}_i-{\mathbf x}_j|)$, where ${\mathbf x}_i$ are the
positions of fermions], the successive terms of the expansion
(\ref{eqn:OmegaPertExpansion}) would be (ultraviolet) finite. If
$\hat V_{\rm int}$ is the local interaction (\ref{eqn:Vint}) of the effective
theory, the successive terms of the expansion (\ref{eqn:OmegaPertExpansion})
involve ultraviolet divergences and have to be regularized. As in our previous
works we will employ for this purpose the cutoff $\Lambda$ on the wave vectors
of virtual particles. Finite (in the limit $\Lambda\rightarrow\infty$)
contributions to the potential $\Omega$ (a physical quantity) are then
obtained by systematically expressing the (bare) couplings of $\hat V_{\rm int}$
in terms of other measurable (physical) quantities. As it is customary, and in
line (at least when the gas is very diluted) with the physical intuition that
properties of the gas are mainly determined by elastic two-body collisions of
its constituents, one expresses the couplings $C_0$, $C_2$, etc. of
(\ref{eqn:Vint}) in terms of the measurable quantities related to such a
scattering process, namely in terms of the scattering lengths $a_0$, $a_1$,
effective ranges $r_0$, etc. \cite{HamFur00}. In this work we will only need
to express the coupling $C_0$ in this way; the relevant formula obtained by
matching the amplitude of the elastic fermion-fermion scattering computed
perturbatively using the first term of the interaction (\ref{eqn:Vint})
onto the general form of the same amplitude parameterized by $a_0$,
$a_1,\dots$ and $r_0,\dots,$ reads
\cite{CHWO1,CHWO3,WeDrSch}
\begin{eqnarray}
  C_0={4\pi\hbar^2\over m_f}~\!a_0\left(1+{2\over\pi}~\!a_0\Lambda
  +{4\over\pi^2}~\!a^2_0\Lambda^2\dots\right)
 \equiv C_0^{\rm R}\left(1+{2\over\pi}~\!a_0\Lambda
  +{4\over\pi^2}~\!a^2_0\Lambda^2\dots\right).\label{C0intermsofCORen}
\end{eqnarray}
\vskip0.1cm

From the thermodynamic point of view much more convenient to work with than
the potential $\Omega$ is the free energy $F$ which canonically depends on the
variables $T$, $V$ and the particle numbers $N_\pm$ which are given by
the derivatives
\begin{eqnarray}
  N_\pm=-(\partial\Omega/\partial\mu_\pm)_{T,V}~\!.\label{eqn:NsFromOmega}
\end{eqnarray}
In principle, in each order of the expansion (\ref{eqn:OmegaPertExpansion})
to construct the free energy one should invert the relations
(\ref{eqn:NsFromOmega}) 
to obtain the chemical potentials as functions of the particle numbers $N_+$
and $N_-$ and insert these in the formula $F=\Omega+\mu_+N_++\mu_-N_-$. Thus
the values of the chemical potentials $\mu_+$ and $\mu_-$ change with each
successive order of the expansion and the procedure of constructing the free
energy looks at first sight rather cumbersome. It turns out, however, that in
the systematic expansion this procedure simplifies considerably: it amounts in
effect to using the chemical potentials $\mu_\pm^{(0)}$ determined by inverting
the formula (\ref{eqn:NsFromOmega}) with $\Omega$ replaced by $\Omega^{(0)}$
given by (\ref{eqn:Omega0Textbook}) and omitting
in the expansion (\ref{eqn:OmegaPertExpansion}) those diagrams which give
vanishing contribution in computing the corrections $\Delta E$ to the ground
state energy $E=E^{(0)}+\Delta E$ of the system of interacting particles using
the ordinary Dyson expansion of the formula
\cite{HamFur00,WeDrSch,CHWO1,CHWO3,CHWO4} ($T$ stands here for time, not for
the temperature, and ${\rm T}_t$ means the ordinary chronological ordering)
\begin{eqnarray}
  \Delta E=\lim_{T\rightarrow\infty}{i\hbar\over T}~\!
  \langle0|{\rm T}_t\exp\!\left({1\over i\hbar}\!\int_{-T/2}^{T/2}\!dt~\!
  V_{\rm int}^I(t)\right)\!|0\rangle~\!,\label{eqn:CorrectionsToE}
\end{eqnarray}
in which $|0\rangle$ is the ground state of the noninteracting system of
$N=N_++N_-$ fermions. In
the case of the interaction proportional to $C_0$ this has been explicitly
demonstrated in \cite{CHGR} up to the third order of the perturbative expansion.
This prescription is obviously consistent with the fact that in the zero
temperature limit the corrections to the free energy obtained from the
thermodynamic expansion should go over into the corresponding corrections to
the ground state energy given by (\ref{eqn:CorrectionsToE}).
\vskip0.1cm

The interaction of the system of spin $1/2$ fermions with the external magnetic
field ${\cal H}$ represented by the operator (the magnetic moment of the
fermions is here included in ${\cal H}$)
\begin{eqnarray}
  \hat V_{\rm int}^{({\cal H})}=-{\cal H}\int_V\!d^3{\mathbf x}\left(
  \hat\psi^\dagger_+\hat\psi_+-\hat\psi^\dagger_-\hat\psi_-\right),
\end{eqnarray}
can also be easily taken into account in this formalism by including it in the
free Hamiltonian $\hat H_0$ which amounts to shifting the
chemical potentials $\mu_\pm\rightarrow\tilde\mu_\pm\equiv\mu_\pm\pm{\cal H}$
in $\hat K_0$ given by (\ref{eqn:KandK0}). The free energy is then given
as the series
\begin{eqnarray}
  F(T,V,{\cal H},N_+,N_-)=F^{(0)}+F^{(1)}+F^{(2)}+\dots,
\end{eqnarray}
in which
\begin{eqnarray}
F^{(0)}(T,V,{\cal H},N_+,N_-)=\Omega^{(0)}(T,N,\tilde\mu_+^{(0)},\tilde\mu_-^{(0)})
+(\tilde\mu_+^{(0)}-{\cal H})N_++(\tilde\mu_+^{(0)}+{\cal H})N_-~\!,
\label{eqn:F(0)Term}
\end{eqnarray}
and $F^{(N)}(T,V,{\cal H},N_+,N_-)
=\Omega^{(N)}(T,N,\tilde\mu_+^{(0)},\tilde\mu_-^{(0)})$ for $N=1,2,\dots,$  
where, as explained above, in computing $\Omega^{(1)}$, $\Omega^{(2)}$ etc.
one should take into account only those diagrams of the expansion
(\ref{eqn:OmegaPertExpansion}) which give nonzero contributions to $\Delta E$.
The (shifted) chemical potentials $\tilde\mu_+^{(0)}$, $\tilde\mu_-^{(0)}$
are given by
\begin{eqnarray}
\tilde\nu_\pm^{(0)}\equiv\tilde\mu_\pm^{(0)}/k_{\rm B}T=
f^{-1}\!\left((1\pm P)\left({\varepsilon_{\rm F}(n)\over
k_{\rm B}T}\right)^{3/2}\right),\label{eqn:nu(0)Determination}
\end{eqnarray}
where $\varepsilon_{\rm F}(n)\equiv\hbar^2k_{\rm F}^2/2m_f$ and
$f^{-1}(x)$ is the inverse of the monotonic function (mapping
${\mathbb R}$ onto ${\mathbb R}_+$) defined by the integral
\begin{eqnarray}
f(\nu)={3\over2}\!\int_0^\infty\!d\xi~\!{\xi^{1/2}\over1+e^{\xi-\nu}}~\!.
\end{eqnarray}

If the computation of $F$ is restricted to the order $C^{\rm R}_0$ (i.e. to the
order $k_{\rm F}a_0$) correction given, as follows from the formulated
prescription and the result (\ref{eqn:Omega(1)}) and (\ref{eqn:G(0,0)}), by
\begin{eqnarray}
  F^{(1)}=VC_0(N_+/V)(N_-/V)~\!,\label{eqn:F(1)}
\end{eqnarray}
the condition for the minimum of $F$ with respect to $N_+$ and $N_-$ (at fixed
$N_++N_-=N$) which determines the system's polarization $P$ takes the form
($t\equiv k_{\rm B}T/\varepsilon_{\rm F}$, $h\equiv{\cal H}/\varepsilon_{\rm F}$)
\begin{eqnarray}
  {8\over3\pi}~\!(k_{\rm F}a_0)~\!P+2h=t
  \left[f^{-1}\!\left({1+P\over t^{3/2}}\right)-
    f^{-1}\!\left({1-P\over t^{3/2}}\right)\right].\label{eqn:meanFieldCond}
\end{eqnarray}
If the asymptotic expansion
\begin{eqnarray}
  f^{-1}(x)=x^{2/3}\left[1-(\pi^2/12)x^{-4/3}-(\pi^4/80)x^{-8/3}
    -(247\pi^6/25920)x^{-4}+\dots\right],
\end{eqnarray}
valid for $x\gg1$ [obtained by inverting the Sommerfeld expansion \cite{LL}
of the function $f(\nu)$] is used, the formula (\ref{eqn:meanFieldCond})
reproduces the textbook \cite{Kesio,Pathria} low temperature equilibrium
condition (equivalent to the condition $\mu_+=\mu_-$) and leads
to the well known prediction that the Stoner phase transition to the ordered
state is continuous with divergent magnetic susceptibility characterized
by the critical exponent $\gamma=1$ and a finite discontinuity of the heat
capacity. (In fact, this continuous character of the transition is accidental:
in the same approximation the transition is of first order in the system
of spin $s>1/2$ fermions and/or if the space dimension is different than 3.)
If the correction $F^{(2)}$ is included, the transition becomes first order,
at least at sufficiently low temperatures \cite{DUMacDO,CHWO1,PECABO1}
in agreement with the arguments given in the past in \cite{BeKiVoj}. However
the computation of the complete order $(k_{\rm F}a_0)^3$ correction to the
ground state energy $E$ performed in \cite{CHWO3,CHWO4,PECABO2} showed that
the first order character of the transition at zero temperature, very clear
in the second order approximation, is significantly weakened, at least as
long as the contributions proportional to $k^3_F a^2_0 r_0$ and $(k_F a_1)^3$
(which, if the underlying interaction potential $V_{\rm pot}$ is ``natural'',
i.e. if all $a_\ell$, $r_\ell$, etc. are of the same order of magnitude,
are of the same, $(k_{\rm F}R)^3$, order as the $(k_{\rm F}a_0)^3$ correction)
are not taken into account \cite{PECABO2}. Below we extend the
existing computations in two ways: we compute the complete (proportional
to $(C_0^{\rm R})^3$, i.e. to $(k_{\rm F}a_0)^3$) temperature dependent
third order corrections to the free energy $F$ and, moreover, we show how to
perform the resummation of two infinite subsets of temperature dependent
corrections to $F$ of which the first one is the finite temperature
generalization of the subset of diagrams taken into account in (the last
section of) ref. \cite{He1}.
\vskip0.5cm

\noindent{\large\bf 3. The order $(k_{\rm F}a_0)^2$ and the
  order $(k_{\rm F}a_0)^3$ particle-particle corrections to $F$}
\vskip0.3cm

\begin{figure}[]
\begin{center}
\includegraphics[width = 0.9\textwidth]{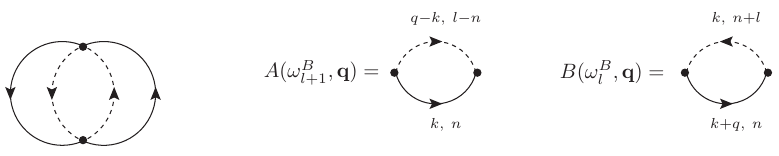}
\end{center}
\caption{The order $C_0^2$ diagram contributing to the thermodynamic potential
  $F$ (or to the ground-state energy $E$) of the gas of spin $1/2$ fermions
  and two ``elementary'' one-loop diagrams ($A$- and $B$-blocks)
  out of which the second order and those higher order (in the $C_0$
  coupling) contributions that are taken into account in this work
  are composed. Solid and dashed lines denote propagators of fermions
  with the spin projections $+$ and $-$, respectively.}
\label{fig:ElementaryLoops}
\end{figure}

\noindent We begin by recalling  the computation of the
order $(C_0^{\rm R})^2$ term $F^{(2)}$ performed in Ref. \cite{CHGR}. In agreement
with the formulated prescription it is given by
the single Feynman diagram shown in figure \ref{fig:ElementaryLoops}. The
corresponding analytical expression can be obtained by convoluting either two
$A$-blocks or two $B$-blocks shown in the same figure:
\begin{eqnarray}
  F^{(2)}=-{1\over2}~\!C_0^2V~\!{1\over\beta}\sum_{l\in\mathbb{Z}}\!
  \int_{\mathbf q}[A(\omega_l^B,\mathbf{q})]^2
  =-{1\over2}~\!C_0^2V~\!{1\over\beta}\sum_{l\in\mathbb{Z}}\!
\int_{\mathbf q}[B(\omega_l^B,\mathbf{q})]^2.\label{eqn:F2InTermsOFAblocks}
\end{eqnarray}
To make the formulas resulting from convoluting $A$-blocks more transparent
it will be convenient to introduce the following notation:
\begin{eqnarray}
&&N_{--}^{\mathbf p}\equiv n_+({\mathbf p})~\!n_-({\mathbf q}-{\mathbf p})~\!,
\nonumber\\
&&N_{++}^{\mathbf p}\equiv[1-n_+({\mathbf p})]~\!
[1-n_-({\mathbf q}-{\mathbf p})]~\!,\nonumber\\
&&n_\pm({\mathbf p})
=\left[1+\exp\{\beta(\varepsilon_{\mathbf p}-\tilde\mu_\pm^{(0)})\}
  \right]^{-1},\label{eqn:Defs}\\
&&\{{\mathbf p}\}\equiv n_+({\mathbf p})+n_-({\mathbf q}-{\mathbf p})-1~\!,
\nonumber\\
&&[{\mathbf p}]\equiv\varepsilon_{\mathbf p}-\tilde\mu_+^{(0)}
  +\varepsilon_{{\mathbf q}-{\mathbf p}}-\tilde\mu_-^{(0)}~\!.\nonumber
\end{eqnarray}
At zero temperature $N_{--}^{\mathbf p}$  and $N_{++}^{\mathbf p}$ reduce
respectively to
$\theta(p_{{\rm F}+}-|{\mathbf p}|)\theta(p_{{\rm F}-}-|{\mathbf q}-{\mathbf p}|)$
and $\theta(|{\mathbf p}|-p_{{\rm F}+})
\theta(|{\mathbf q}-{\mathbf p}|-p_{{\rm F}-})$,
hence the subscripts $--$ and $++$. It is also easy to check that
\begin{eqnarray}
  \{{\mathbf p}\}=N_{--}^{\mathbf p}-N_{++}^{\mathbf p}
  =N_{--}^{\mathbf p}\left(1-e^{\beta[{\mathbf p}]}\right).\label{eqn:Ids}
\end{eqnarray}
The form of the distribution functions $n_\pm({\mathbf p})$ plays the
role only in the second of these two identities.

In the introduced notation the $A$-block (obtained in Ref. \cite{CHGR}) takes the
form
\begin{eqnarray}
  A(\omega_l^B,\mathbf{q})=\int_{\mathbf p}\!{\{{\mathbf p}\}\over
    i\omega_l^B-[{\mathbf p}]}~\!,\label{eqn:AblockExplicit}
\end{eqnarray}
After the sum in (\ref{eqn:F2InTermsOFAblocks}) over the bosonic Matsubara
frequencies $\omega^B_l=(\pi/\beta)l$ is performed
using the standard formulas \cite{FetWal,CHGR} one gets
\begin{eqnarray}
  {F^{(2)}\over V}=-{1\over2}~\!C_0^2\!\int_{\mathbf q}\!\int_{{\mathbf p}_1}\!
  \int_{{\mathbf p}_2}\!{\{{\mathbf p}_1\}\{{\mathbf p}_2\}
\over[{\mathbf p}_1]-[{\mathbf p}_2]}
  \left({1\over1-e^{\beta[{\mathbf p}_1]}}-{1\over1-e^{\beta[{\mathbf p}_2]}}\right).
  \label{eqn:F2SymmetricForm}
\end{eqnarray}
Since the two terms are formally identical (after making in the integrals in
one of the terms the interchange ${\mathbf p}_1\leftrightarrow{\mathbf p}_2$),
one arrives, using (\ref{eqn:Ids}), at the final form of $F^{(2)}/V$:
\begin{eqnarray}
{F^{(2)}\over V}=C_0^2\!\int_{\mathbf q}\!\int_{{\mathbf p}_1}\!\int_{{\mathbf p}_2}\!
N_{--}^{{\mathbf p}_1}~\!{1\over[{\mathbf p}_1]-[{\mathbf p}_2]}
-C_0^2\!\int_{\mathbf q}\!\int_{{\mathbf p}_1}\!\int_{{\mathbf p}_2}\!
N_{--}^{{\mathbf p}_1}~\!
{\{{\mathbf p}_2\}^{\rm sub}\over[{\mathbf p}_1]-[{\mathbf p}_2]}~\!,
\label{eqn:F(2)inTermsOfC0}
\end{eqnarray}
in which
$\{{\mathbf p}\}^{\rm sub}\equiv\{{\mathbf p}\}+1
= n_+({\mathbf p})+n_-({\mathbf q}-{\mathbf p})$
and the integrals should be understood in the principal value sense. Notice
that the denominators $[{\mathbf p}_1]-[{\mathbf p}_2]$ do not depend on the
chemical potentials. This profiting from the symmetry of the two terms of
(\ref{eqn:F2SymmetricForm}), seemingly not problematic, has indeed
no consequences here but, as will be shown, in higher orders if applied
blindly would lead to incorrect results.

The first term in (\ref{eqn:F(2)inTermsOfC0}) is divergent. The change of the
variables
${\mathbf q}=2{\mathbf s}$, ${\mathbf p}_1={\mathbf s}-{\mathbf t}_1$, and
${\mathbf p}_2={\mathbf s}-{\mathbf t}_2$ (the Jacobian equals 8) makes the
innermost integral elementary and allows to write it in the form 
\begin{eqnarray}
  {16\pi^2\hbar^4\over m^2_f}~\!a^2_0
  \left(1+{4\over\pi}~\!a_0\Lambda+\dots\right)\!
  \int_{\mathbf s}\!\int_{{\mathbf t}_1}\!
  8~\!n_+({\mathbf s}-{\mathbf t}_1)~\!n_-({\mathbf s}+{\mathbf t}_1)~\!
{m_f\over2\pi^2\hbar^2}
  \left(-\Lambda+{{\mathbf t}^2_1\over\Lambda}+\dots\right),\nonumber
\end{eqnarray}
after  using (\ref{C0intermsofCORen}).

Expressing $C_0$ similarly in the second term of  the formula
(\ref{eqn:F(2)inTermsOfC0}) and in $F^{(1)}$ given by (\ref{eqn:F(1)}),
one finds that the divergent terms of order $a_0^2\Lambda$ cancel out and
\begin{eqnarray}
{F^{(1)}+F^{(2)}\over V}=C_0^{\rm R}\!\int_{{\mathbf k}_1}\!\int_{{\mathbf k}_2}\!
n_+({\mathbf k}_1)~\!n_-({\mathbf k}_2) -(C_0^{\rm R})^2\int_{\mathbf q}\!
\int_{{\mathbf p}_1}\!\int_{{\mathbf p}_2}\!N_{--}^{{\mathbf p}_1}~\!
{\{{\mathbf p}_2\}^{\rm sub}\over[{\mathbf p}_1]-[{\mathbf p}_2]}\nonumber\\
-{16\hbar^2\over\pi m_f}~\!a_0^3\!\int_{\mathbf s}\!\int_{{\mathbf t}_1}
8~\!n_+({\mathbf s}-{\mathbf t}_1)~\!n_-({\mathbf s}+{\mathbf t}_1)~\!
(\Lambda^2-2{\mathbf t}_1^2)\phantom{aaaaaaaaa}~\!\label{eqn:F1AndF2}\\
-{64\pi\hbar^4\over m_f^2}~\!a_0^3\Lambda\!
\int_{\mathbf q}\!\int_{{\mathbf p}_1}\!\int_{{\mathbf p}_2}\!\
N_{--}^{{\mathbf p}_1}~\!
{\{{\mathbf p}_2\}^{\rm sub}
  \over[{\mathbf p}_1]-[{\mathbf p}_2]}+{\cal O}(1/\Lambda)~\!.
\phantom{aaaaaaaa}~\!\nonumber
\end{eqnarray}
The first two terms constitute the complete, finite contribution to $F/V$ up
to the order $(C_0^{\rm R})^2$; the remaining terms are formally of higher order
and can be considered only after including other third and higher order
contributions.

In Ref. \cite{CHGR} it has been found that it is convenient to evaluate
(the finite part of) $F^{(2)}/V$ by substituting ${\mathbf p}_1={\mathbf k}_1$,
 and ${\mathbf q}={\mathbf k}_1+{\mathbf k}_2$, ${\mathbf p}_2={\mathbf p}$
(the Jacobian is 1), replacing (by another change of the integration variable)
$n_-({\mathbf k}_1+{\mathbf k}_2-{\mathbf p})$
with $n_-({\mathbf p})$ and then performing explicitly the integral over the
cosine of the angle between ${\mathbf p}$ and ${\mathbf k}_1+{\mathbf k}_2$.
This allows to represent the order $(k_{\rm F}a_0)^2$ contribution to
$F$ in the form
\begin{eqnarray}
  {F^{(2)}\over V}=C_0^{\rm R}\!\int_{{\mathbf k}_1}\!\int_{{\mathbf k}_2}\!
  n_+({\mathbf k}_1)~\!n_-({\mathbf k}_2)~\!L({\mathbf k}_1,{\mathbf k}_2)~\!.
  \label{eqn:F(2)final}
\end{eqnarray}
The (dimensionless) function $L({\mathbf k}_1,{\mathbf k}_2)$ is given
by the single integral
\begin{eqnarray}
  L({\mathbf k}_1,{\mathbf k}_2)=
  -{C_0^{\rm R}m_f\over(2\pi)^2\hbar^2|{\mathbf k}_1+{\mathbf k}_2|}
  \!\int_0^\infty\!dp~\!p~\![n_+(p)+n_-(p)]\ln\!
  \left|{(p-\Delta_+)(p-\Delta_-)\over(p+\Delta_+)(p+\Delta_-)}\right|,
  \label{eqn:Lfunction}
\end{eqnarray}
in which
\begin{eqnarray}
\Delta_\pm={1\over2}|{\mathbf k}_1+{\mathbf k}_2|\pm{1\over2}
|{\mathbf k}_1-{\mathbf k}_2|~\!.
\end{eqnarray}
In \cite{CHGR} we have checked that in the zero temperature limit, in which the
Fermi distribution functions $n_+({\mathbf p})$ and $n_-({\mathbf p})$ are
replaced by the step functions $\theta(p_{{\rm F}+}-|{\mathbf p}|)$ and
$\theta(p_{{\rm F}-}-|{\mathbf p}|)$, this formula reproduces
numerically the second order
correction to the ground state energy computed first (analytically) by
Kanno \cite{KANNO} and then recovered (semi-analytically) in several
works (e.g. in \cite{CHWO1,PECABO1}) for all values of the polarization $P$.
We have also analyzed the free energy $F$ with the corrections $F^{(1)}$ and
$F^{(2)}$ included and recovered, up to uncertainties following from the finite
precision of the (rather complicated) numerical evaluation of the relevant
multiple integrals the characteristics of the phase transition to the
ordered state (for temperatures $T<\varepsilon_{\rm F}/k_{\rm B}$) first obtained
in \cite{DUMacDO}.
\vskip0.1cm

\begin{figure}[]
\begin{center}
\includegraphics[width = 0.6 \textwidth]{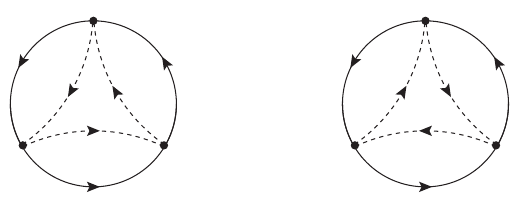}
\end{center}
\caption{The particle-particle and particle-hole diagrams contributing
  in the order $C^3_0$ to the thermodynamic potential $F$ (or to the
  ground-state energy $E$).}
\label{fig:C0cubeMercedes}
\end{figure}

If only the interaction term proportional to $C_0$ in (\ref{eqn:Vint}) is taken
into account, there are two Feynman diagrams contributing to the free energy
$F$ in the third order. The first one, shown in the left panel of Fig.
\ref{fig:C0cubeMercedes}, is termed the particle-particle ring diagram. Its
contribution $F^{(3)pp}$ is given by the convolution of three $A$-blocks
\begin{eqnarray}
  {F^{(3)pp}\over V}={1\over3}~\!C_0^3~\!{1\over\beta}\sum_l
  \int_{\mathbf q}\![A(\omega_l^{\rm B},\mathbf{q})]^3~\!.\label{eqn:F(3)pp}
\end{eqnarray}
After decomposing the product of three $A$-blocks into simple fractions,
performing the summation over the bosonic Matsubara frequencies
$\omega_l^{\rm B}$ and then using the identities (\ref{eqn:Ids}) one arrives at
\begin{eqnarray}
{F^{(3)pp}\over V}={1\over3}~\!C_0^3\!\int_{\mathbf q}\!
\int_{{\mathbf p}_1}\!\int_{{\mathbf p}_2}\!\int_{{\mathbf p}_3}\!
\left(N_{--}^{{\mathbf p}_1}~\!
{\{{\mathbf p}_2\}\over[{\mathbf p}_1]-[{\mathbf p}_2]}~\!
{\{{\mathbf p}_3\}\over[{\mathbf p}_1]-[{\mathbf p}_3]}
+{\rm two~other~terms}\right),\label{eqn:SymmetricFormOfF3pp}
\end{eqnarray}
where ``two other terms'' means the terms in which the role of ${\mathbf p}_1$
is played by ${\mathbf p}_2$ and ${\mathbf p}_3$. It is good at this
point to make  a contact with the contribution of this third order particle-particle
ring diagram to the ground state energy density $E/V$ obtained in Refs.
\cite{CHWO3,CHWO4} (and in Refs. \cite{PECABO2}) to which the expression
(\ref{eqn:SymmetricFormOfF3pp}) should reduce in the zero temperature limit,
that is when the Fermi distribution functions are replaced by the theta
functions. The contribution to $E/V$ of the particle-particle diagram was given in
Refs. \cite{CHWO3,CHWO4}  by two terms (Eq. (21) in Ref. \cite{CHWO4})
whereas here it is given by the single
term (three seemingly identical terms). The equivalence of the two approaches
is ensured by the algebraic, i.e. independent of
the precise forms of $N_{--}^{\mathbf p}$ and $[{\mathbf p}]$ (recall that
$\{{\mathbf p}\}=N_{--}^{\mathbf p}-N_{++}^{\mathbf p}$), identity which results
from the symmetrization
\begin{eqnarray}
N_{--}^{{\mathbf p}_1}~\!
{N_{--}^{\mathbf{p}_2}-N_{++}^{\mathbf{p}_2}\over[\mathbf{p}_1]-[\mathbf{p}_2]}~\!
{N_{--}^{\mathbf{p}_3}-N_{++}^{\mathbf{p}_3}\over[\mathbf{p}_1]-[\mathbf{p}_3]}
+{\rm two~other~terms}\phantom{aaaaaaaaaaaaaaaaaaaaaaaaaaaaa}
\label{eqn:AlgId}\\
=\left(N_{--}^{\mathbf{p}_1}~\!
{N_{++}^{\mathbf{p}_2}\over[\mathbf{p}_1]-[\mathbf{p}_2]}~\!
{N_{++}^{\mathbf{p}_3}\over[\mathbf{p}_1]-[\mathbf{p}_3]}
+N_{++}^{\mathbf{p}_1}~\!{N_{--}^{\mathbf{p}_2}\over[\mathbf{p}_1]-[\mathbf{p}_2]}~\!
{N_{--}^{\mathbf{p}_3}\over[\mathbf{p}_1]-[\mathbf{p}_3]}\right)
+{\rm two~other~terms}.\nonumber
\end{eqnarray}
After using the symmetry,
i.e. taking only the content of the bracket and multiplying it by 3,
it allows to rewrite the expression for $F^{(3)pp}/V$ as the sum of two
terms which in the $T=0$ limit precisely reduce to the two terms, $G_1$
and $G_2$, which in Refs. \cite{CHWO3,CHWO4} contributed to $E/V$.

Naively, as all three terms of (\ref{eqn:SymmetricFormOfF3pp}) seem also
identical, one is tempted to compute only one of them and multiply the result
by three. $F^{(3)pp}/V$ would be in this way given by a single fourfold
integral. This, as we have found, would lead to an incorrect result which in
the zero temperature limit would not agree with the contribution of the
particle-particle diagram to $E/V$ (this is precisely the problem that
did not allow us to immediately extend the computation reported in Ref. \cite{CHGR}).

To understand the problem it is instructive to consider the triple integral
\begin{eqnarray}
 \int_0^1\!dx\!\int_0^1\!dy\!\int_0^1\!dz\left({1\over(x-y)(x-z)}
  +{1\over(y-x)(y-z)}+{1\over(z-x)(z-y)}\right).\label{eqn:Integral}
\end{eqnarray}
The integrand is algebraically zero and the result of the integration should
be zero too. Yet the integrand has (spurious) singularities and the integrals
in (\ref{eqn:Integral}), similarly as the ones encountered in the computation
of $F^{(3)}$, should be understood in the principal value
sense. If one naively says that the integrals of the three terms are equal
and evaluates only one of them (multiplying it by 3) one will get
\begin{eqnarray}
 3\int_0^1\!dx\!\left({\rm P}\!\int_0^1\!dy~\!{1\over x-y}\right)^2
 =3\int_0^1\!dx~\!\ln^2{1-x\over x}=3~\!{\pi^2\over3}.\nonumber
\end{eqnarray}
The correct result (zero) is obtained if one first regularizes the integrand
of (\ref{eqn:Integral}) by
setting $x\rightarrow x+i\epsilon$, $y\rightarrow y+2i\epsilon$,
$z\rightarrow z+3i\epsilon$ (the sign of $\epsilon$ is irrelevant;
the integrand is still algebraically zero
but its singularities are now off the integration axes). It is then
straightforward to find that the application of the Sochocki formula
$1/(x\pm i0)=P(1/x)\mp i\pi\delta(x)$ to the regularized integral
(\ref{eqn:Integral}) leads to
(the terms linear in the Dirac deltas neatly cancel out)
\begin{eqnarray}
 3\int_0^1\!dx\!\left({\rm P}\!\int_0^1\!dy~\!{1\over x-y}\right)^2
 +\int_0^1\!dx\!\left(i\pi\!\int_0^1\!dy~\!\delta(x-y)\right)^2=0~\!.\nonumber
\end{eqnarray}
If the same procedure is applied to (\ref{eqn:SymmetricFormOfF3pp}) one gets
\begin{eqnarray}
  {F^{(3)pp}\over V}=C_0^3\!\int_{\mathbf q}\!\int_{{\mathbf p}_1}\!~
  N_{--}^{{\mathbf p}_1}\!\left[\left({\rm P}\!\int_{{\mathbf p}_2}~\!
  {\{{\mathbf p}_2\}\over[{\mathbf p}_1]-[{\mathbf p}_2]}\right)^2
  +{1\over3}\left(i\pi\!\int_{{\mathbf p}_2}\!\{{\mathbf p}_2\}~\!
  \delta([{\mathbf p}_1]-[{\mathbf p}_2])\right)^2\right].
  \label{eqn:F3ppCorrect}
\end{eqnarray}

One can now check that in the sum $F^{(1)}+F^{(2)}+F^{(3)pp}$ all the divergences
[up to the order $(k_{\rm F}a_0)^3$] cancel out (as will be seen, the
contribution $F^{(3)ph}$ of the other diagram of Figure \ref{fig:C0cubeMercedes},
which completes the order $(k_{\rm F}a_0)^3$ contribution to $F^{(3)}$, is finite;
this also follows from the computations of the
order $(k_{\rm F}a_0)^3$ corrections to $E/V$ performed in Refs. \cite{CHWO3,CHWO4}).
Writing $\{{\mathbf p}_i\}=-1+\{{\mathbf p}_i\}^{\rm sub}$ in the
first term in the square bracket in (\ref{eqn:F3ppCorrect}) allows to single
out the divergent part of $F^{(3)}/V$. It is given by (to this order one can
set in (\ref{eqn:F3ppCorrect}) $C_0=C_0^{\rm R}$; we also suppress the symbol
$P$ of the principal value)
\begin{eqnarray}
  {F^{(3)pp}_{\rm div}\over V}={64\pi^3\hbar^6\over m_f^3}~\!a_0^3
  \!\int_{\mathbf q}\!\int_{{\mathbf p}_1}
  N_{--}^{{\mathbf p}_1}\!\left(\int_{{\mathbf p}_2}~\!
  {1\over[{\mathbf p}_1]-[{\mathbf p}_2]}\right)^2
  \phantom{aaaaaaaaaaa}\nonumber\\
  -2~\!{64\pi^3\hbar^6\over m_f^3}~\!a_0^3\!\int_{\mathbf q}\!\int_{{\mathbf p}_1}
  N_{--}^{{\mathbf p}_1}\!\int_{{\mathbf p}_2}~\!
  {\{{\mathbf p}_2\}^{\rm sub}\over[{\mathbf p}_1]-[{\mathbf p}_2]}
  \int_{{\mathbf p}_3}~\!{1\over[{\mathbf p}_1]-[{\mathbf p}_3]}~\!.
\end{eqnarray}
Making now in the first term the change of the variables
${\mathbf q}=2{\mathbf s}$, ${\mathbf p}_1={\mathbf s}-{\mathbf t}_1$,
${\mathbf p}_2={\mathbf s}-{\mathbf t}_2$ (the Jacobian is 8)
and performing the innermost integral (over $d^3{\mathbf t}_2$) one finds
that it precisely cancels the entire middle line of (\ref{eqn:F1AndF2}).
Moreover, after making similar changes of the variables in the
last line of (\ref{eqn:F1AndF2}) and in the last term of
$F^{(3)pp}_{\rm div}/V$ they too mutually cancel out.
The remaining contribution of the left diagram
of Figure \ref{fig:C0cubeMercedes} is, therefore, given
by (\ref{eqn:F3ppCorrect}) with
$\{{\mathbf p}_2\}$ in the first term (but not in the second one!)
replaced by $\{{\mathbf p}_2\}^{\rm sub}$ and $C_0$ replaced by $C_0^{\rm R}$. 
Making as previously the change ${\mathbf p}_1={\mathbf k}_1$,
${\mathbf q}={\mathbf k}_1+{\mathbf k}_2$, ${\mathbf p}_2={\mathbf p}$
of the integration variables
one can represent the  contribution of the particle-particle 
order $(k_{\rm F}a_0)^3$ diagram to $F/V$ in the form
\begin{eqnarray}
  {F^{(3)pp}_{\rm fin}\over V}=C_0^{\rm R}\!\int_{{\mathbf k}_1}\!\int_{{\mathbf k}_1}\!
  n_+({\mathbf k}_1)~\!n_-({\mathbf k}_2)\left[L^2({\mathbf k}_1,{\mathbf k}_2)
    +{1\over3}\left(iL_\delta({\mathbf k}_1,{\mathbf k}_2)\right)^2\right],
  \label{eqn:F(3)ppfinal}
\end{eqnarray}
where the function $L({\mathbf k}_1,{\mathbf k}_2)$ is given by 
(\ref{eqn:Lfunction}) while the dimensionless function
$L_\delta({\mathbf k}_1,{\mathbf k}_2)$ is given by the finite integral
\begin{eqnarray}
  L_\delta({\mathbf k}_1,{\mathbf k}_2)
  =\pi~\!{C_0^{\rm R}m_f\over(2\pi)^2\hbar^2|{\mathbf k}_1+{\mathbf k}_2|}\!
  \int_{p_{\rm min}}^{p_{\rm max}}\!dp~\!p~\![n_+(p)+n_-(p)-1]~\!,
  \label{eqn:Ldeltafunction}
\end{eqnarray}
in which $p_{\rm min}=|\Delta_-|$ and $p_{\rm max}=\Delta_+$ are determined by
the condition that the zero of the argument of the Dirac delta treated as a
function of the cosine of the angle between ${\mathbf p}$ and
${\mathbf k}_1+{\mathbf k}_2$ lies between $-1$ and $+1$.
Since $\varepsilon_{\mathbf p}$ depends on $p^2$,
the function $L_\delta({\mathbf k}_1,{\mathbf k}_2)$ can, of course, be
obtained in a closed analytic form.
The expression (\ref{eqn:F(3)ppfinal}) is finite (as its superscript indicates)
the ultraviolet convergence\footnote{The singularity introduced by the 
  factor $1/|{\mathbf k}_1+{\mathbf k}_2|^2$ is superficial because
  for ${\mathbf k}_1+{\mathbf k}_2={\mathbf0}$ vanishes also 
  the logarithm in (\ref{eqn:Lfunction}) while in (\ref{eqn:Ldeltafunction})
  $p_{\rm min}=p_{\rm max}$.}
of the integrations being secured by the exponential suppression provided by
the Fermi distribution functions $n_+({\mathbf k}_1)$ and $n_-({\mathbf k}_2)$.
We have also checked that the expression
(\ref{eqn:F(3)ppfinal})
evaluated for $T=0$ (so that the Fermi distribution functions can be replaced
by the step functions) reproduces numerically in the entire range of
polarizations $P$ the contribution to the ground state energy density
of the third order particle-particle diagram 
of Figure \ref{fig:C0cubeMercedes} obtained in Refs. \cite{CHWO3,CHWO4}.
\vskip0.5cm

\noindent{\large\bf 4. Resummation of the contributions of the particle-particle
diagrams}
\vskip0.3cm

\noindent It turns out that the contribution to the free energy $F$ of the
infinite series of Feynman diagrams composed of $N$-fold rings of the
particle-particle $A$-blocks of Figure \ref{fig:ElementaryLoops}
can be summed in a closed form. Consider first the order
$(k_{\rm F}a_0)^N$ term of this series [the
factor $(-1)^{N+1}$ is the same as in (\ref{eqn:OmegaPertExpansion}) - there
are as many rearrangements of the $\hat\psi_+$ operators as of the $\hat\psi_-$
ones; the factor $1/N!$ in (\ref{eqn:OmegaPertExpansion}) is reduced to
$1/N$ as there are $(N-1)!$ identical diagrams]:
\begin{eqnarray}
  {F^{(N)pp}\over V}=(-1)^{N+1}~\!{C_0^N\over N}~\!{1\over\beta}\sum_l
  \int_{\mathbf q}[A(\omega_l^{\rm B},{\mathbf q})]^N~\!.
\end{eqnarray}
Decomposing the product of the integrands of the $N$ $A$-blocks
(\ref{eqn:AblockExplicit}) using the identity
\begin{eqnarray}
\prod_{i=1}^N{1\over x-a_i}=\sum_{n=1}^N\left(\prod_{j\neq n}^N{1\over a_n-a_j}
\right){1\over x-a_n}~\!,
\end{eqnarray}
and performing then the summation over the Matsubara frequencies one gets
the integrand of the $(N+1)$-fold integral in the form
\begin{eqnarray}
  \{{\mathbf p}_1\}\dots\{{\mathbf p}_N\}
  \sum_{n=1}^N\left(\prod_{j\neq n}^N{1\over[{\mathbf p}_n]-[{\mathbf p}_j]}\right)
  {1\over1-e^{\beta[{\mathbf p}_n]}}~\!,
\end{eqnarray}
and finally, after using the relations (\ref{eqn:Ids}),
$F^{(N)pp}/V$ takes the form
\begin{eqnarray}
  {F^{(N)pp}\over V}=(-1)^{N+1}~\!{C_0^N\over N}\!\int_{\mathbf q}\!
  \int_{{\mathbf p}_1}\!\dots\!\int_{{\mathbf p}_N}\sum_{n=1}^N
  N_{--}^{{\mathbf p}_n}\left(\prod_{j\neq n}^N
  { \{ {\mathbf p}_j \}\over[{\mathbf p}_n]-[{\mathbf p}_j]
    +i(n-j)\epsilon}\right),
\end{eqnarray}
in which, in order to regularize the integrals, the substitution
$[{\mathbf p}_l]\rightarrow[{\mathbf p}_l]+il\epsilon$ has been made. Using the 
Sochocki formula this can be then rewritten (assuming that $\epsilon>0$
-- it will
be seen that the result does not depend on the sign of $\epsilon$) in the form
\begin{eqnarray}
{F^{(N)pp}\over V}=(-1)^{N-1}~\!{C_0^N\over N}\!\int_{\mathbf q}
\sum_{n=1}^N\int_{{\mathbf p}_n}\!\!N_{--}^{{\mathbf p}_n}
\left\{\prod_{j=1}^{n-1}\int_{{\mathbf p}_j}\!\left(
{\{{\mathbf p}_j\}\over[{\mathbf p}_n]-[{\mathbf p}_j]}
-i\pi~\!\{{\mathbf p}_j\}~\!\delta([{\mathbf p}_n]-[{\mathbf p}_j])\right)
\right.\nonumber\\
\left.\times\prod_{j=n+1}^N\int_{{\mathbf p}_j}\!\left(
{\{{\mathbf p}_j\}\over[{\mathbf p}_n]-[{\mathbf p}_j]}
+i\pi~\!\{{\mathbf p}_j\}~\!
\delta([{\mathbf p}_n]-[{\mathbf p}_j])\right)\right\},\nonumber
\end{eqnarray}
in which the integrals of the factors
$\{{\mathbf p}_j\}/([{\mathbf p}_n]-[{\mathbf p}_j])$ are understood in the
principal value sense. The experience with the order $(k_{\rm F}a_0)^2$ and
$(k_{\rm F}a_0)^3$ contributions teaches that removing divergences amounts
simply to replacing $\{{\mathbf p}_j\}$ by $\{{\mathbf p}_j\}^{\rm sub}$ in
the first terms of the integrands of the integrals over ${\mathbf p}_j$-s
(but not in the delta-terms) and $C_0^N$ in front by $(C_0^{\rm R})^N$. This
would be obvious had the dimensional regularization been used to handle
ultraviolet divergences --
\begin{figure}
\centerline{\hbox{
\psfig{figure=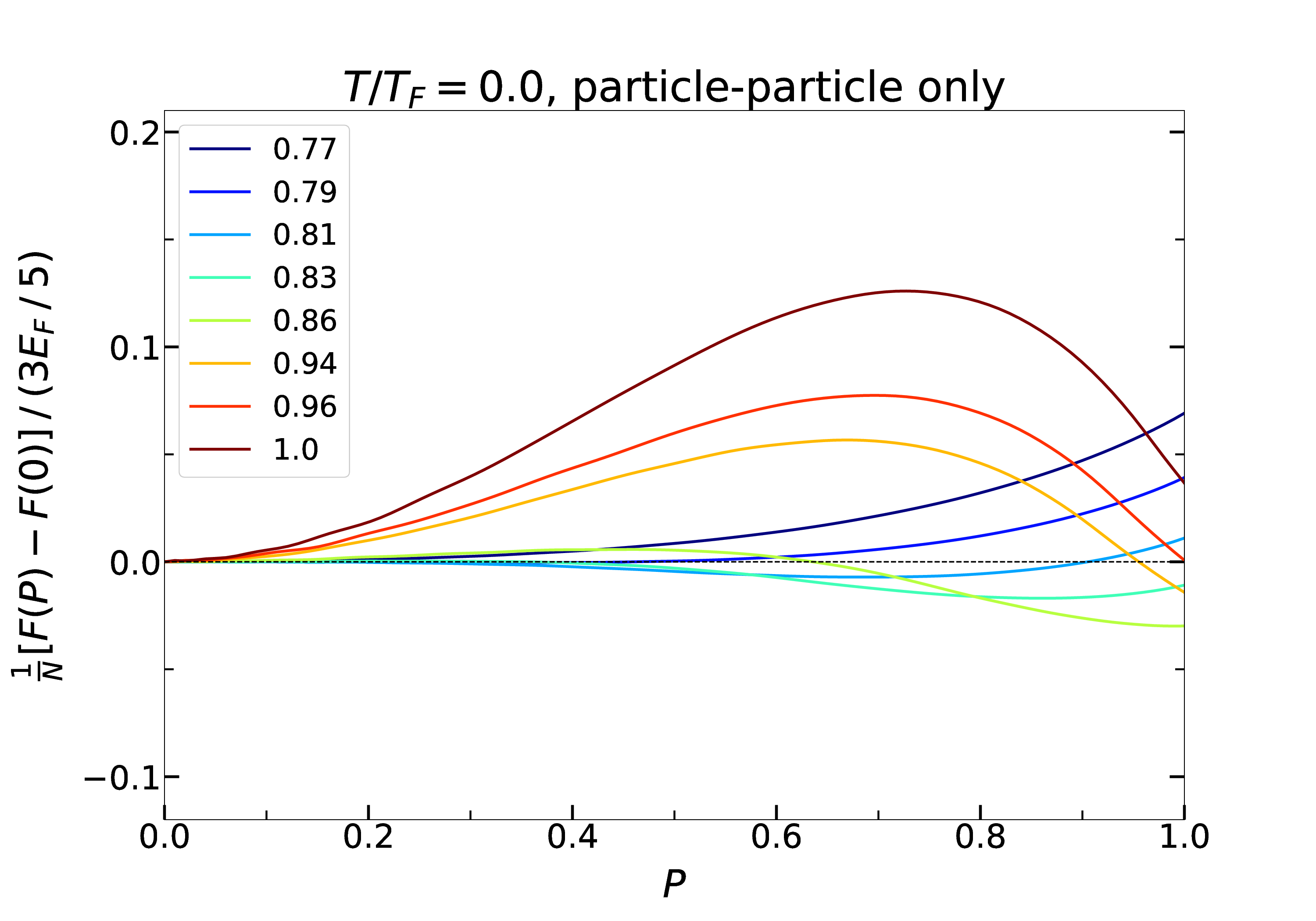,width=9.cm,height=7.0cm} 
\psfig{figure=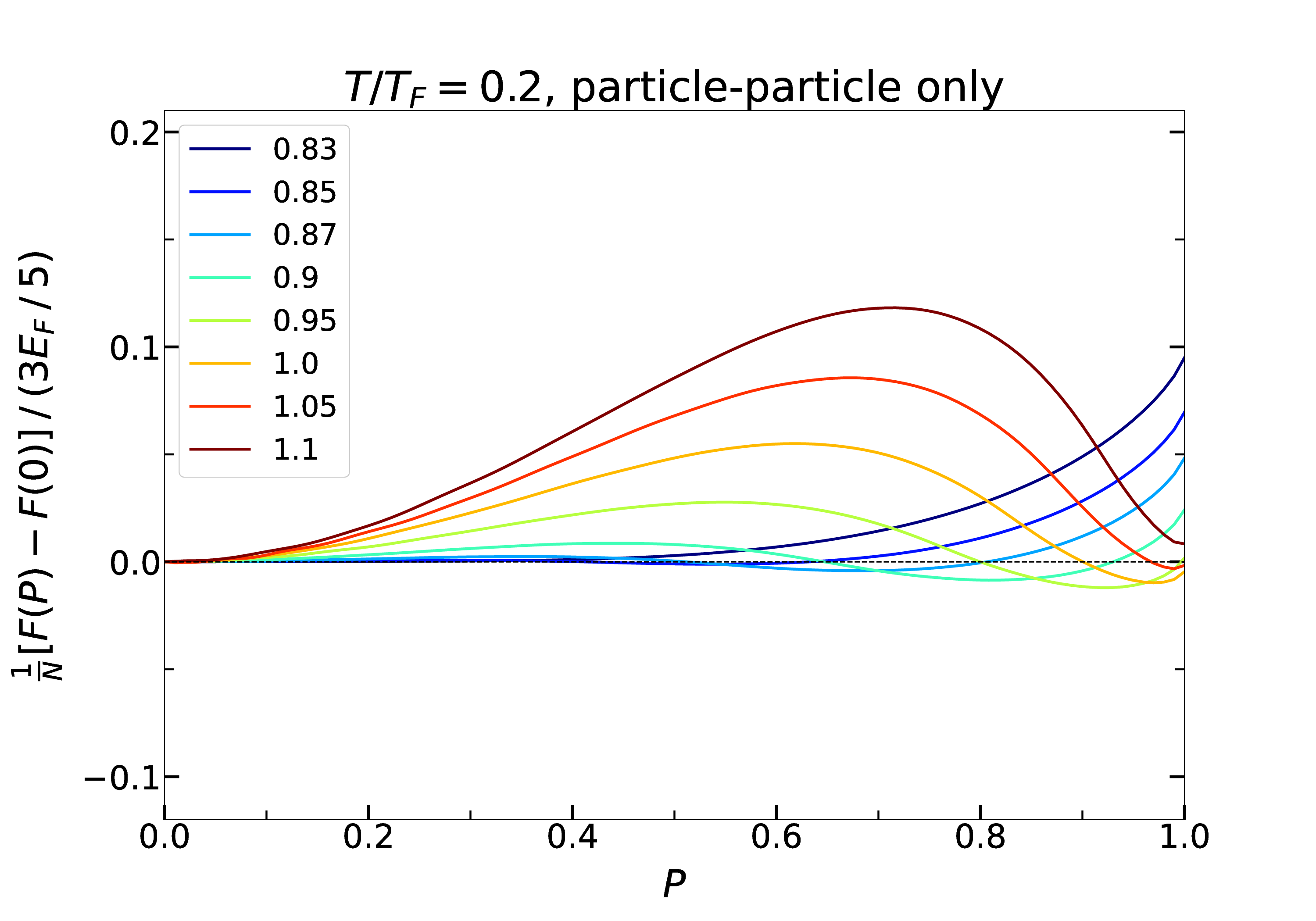,width=9.cm,height=7.0cm} 
}}
\caption{The difference $(F(P)-F(0))^{pp}/N$ (in units
  $(3/5)\varepsilon_{\rm F}$) for $T=0$ and
  $T=0.2~\!T_{\rm F}\equiv0.2~\!\varepsilon_{\rm F}/k_{\rm B}$ as a function of the
  polarization $P=(N_+-N_-)/N$ for different values (indicated in the panes)
  of the gas parameter $k_{\rm F}a_0$.}
\label{fig:FresPPonlyT00and02}
\end{figure}
by definition it
sets the integrals like $\int_{\mathbf p}({\rm const})$ to zero and, as is well
known (see e.g. Ref. \cite{HamFur00}), $C_0=C_0^{\rm R}$  to all orders, if such a
regularization is used.
After the change of the variables ${\mathbf p_n}={\mathbf k}_1$,
and ${\mathbf q}={\mathbf k}_1+{\mathbf k}_2$
the order $(k_{\rm F}a_0)^N$ particle-particle diagram contribution to $F$
can be neatly written in the form
\begin{eqnarray}
  {F^{(N)pp}\over V}={C_0^{\rm R}\over N}\!
  \int_{{\mathbf k}_1}\!\int_{{\mathbf k}_2}\!n_+({\mathbf k}_1)~\!n_-({\mathbf k}_2)
  \sum_{n=1}^N(L+iL_\delta)^{n-1}(L-iL_\delta)^{N-n}~\!.
\end{eqnarray}
Summing the (finite) geometric series then gives
\begin{eqnarray}
  {F^{(N)pp}\over V}={C_0^{\rm R}\over N}\!
  \int_{{\mathbf k}_1}\!\int_{{\mathbf k}_2}\!n_+({\mathbf k}_1)~\!n_-({\mathbf k}_2)
  ~\!{(L+iL_\delta)^N-(L-iL_\delta)^N\over2iL_\delta}~\!.\label{eqn:FNpp}
\end{eqnarray}
This is real and independent of the sign of $L_\delta$ which reflects the fact
that in the prescription allowing to properly handle the $P$-value integrals
the sign of $\epsilon$ is arbitrary; in particular it has nothing to do with
the prescription $+i0^+$ for standard Feynman propagators in the real time
formalism. As can be easily checked, for $N=2$ and $N=3$ (\ref{eqn:FNpp})
reproduces the results of (\ref{eqn:F(2)final}) and (\ref{eqn:F(3)ppfinal}),
respectively. It is also clear that for $N=1$ the mean field result
(\ref{eqn:F(1)}) is recovered.
Finally, summation over $N$ can also be done\footnote{We use the formula
  arctan$~\!t=(1/2i)\ln[(1+it)/(1-it)]$ and the expansion of the logarithm
  in powers of $t$.}
and leads to the expression
\begin{eqnarray}
  {F^{pp}\over V}=C_0^{\rm R}\int_{{\mathbf k}_1}\!\int_{{\mathbf k}_2}\!
  n_+({\mathbf k}_1)~\!n_-({\mathbf k}_2)
  ~\!{{\rm arctan}(L_\delta/(1-L))\over L_\delta}~\!.\label{eqn:FppSummed}
\end{eqnarray}
The zero temperature analog of this formula (i.e. representing the contribution
of the particle-particle ring diagrams to the gas energy density
$E/V$) has been for $P=0$ first given by Kaiser \cite{KAJZERKA1} who in
deriving it relied on combinatoric arguments. The formula that  is
the zero temperature analog of (\ref{eqn:FppSummed}) for arbitrary polarization
$P$ has then been written (by invoking the Kaiser's reasoning)
down in Refs. \cite{He1,He3} (see also Ref. \cite{KAJZERKA2}).

\begin{figure}
\centerline{\hbox{
\psfig{figure=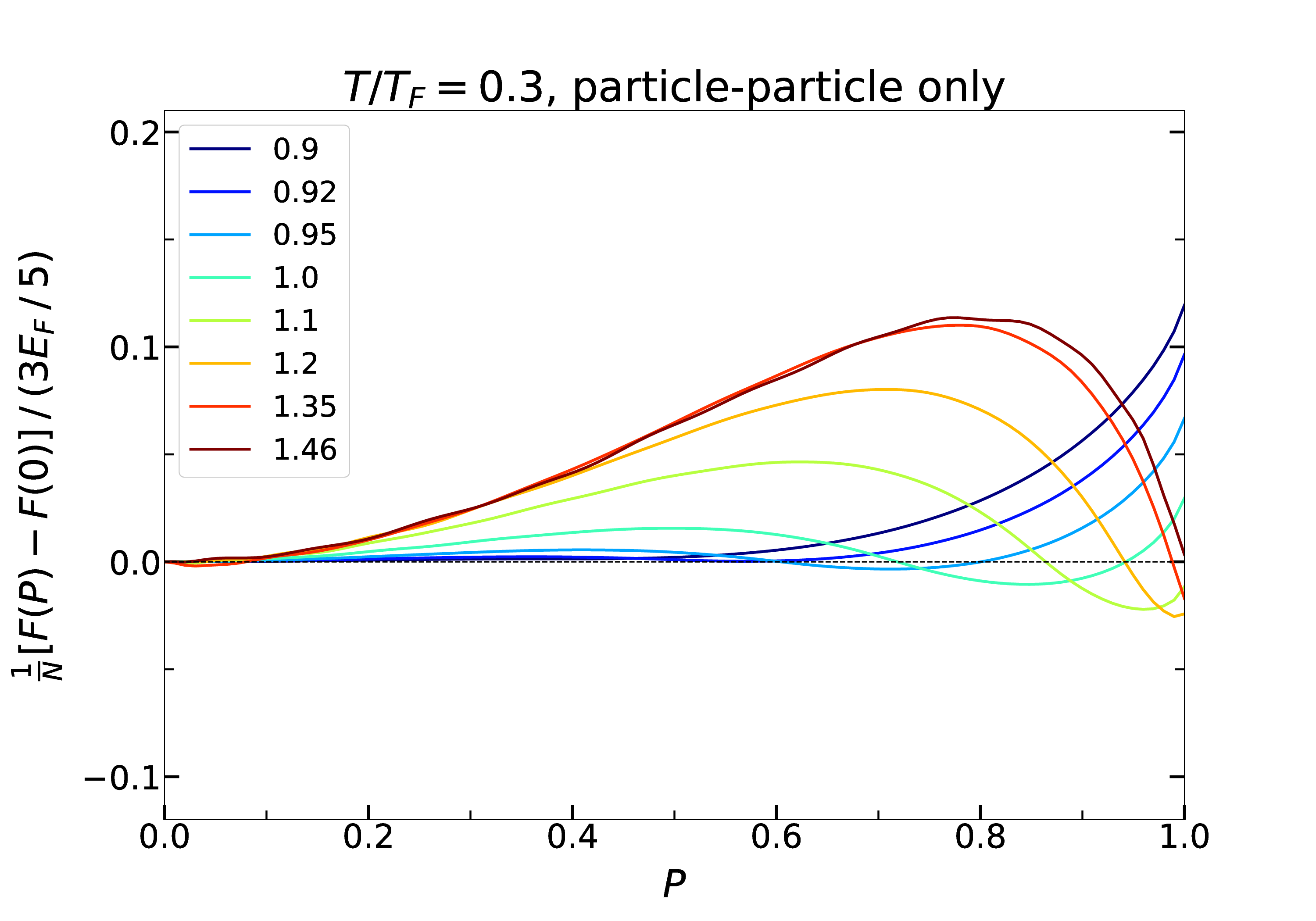,width=9.cm,height=7.0cm} 
\psfig{figure=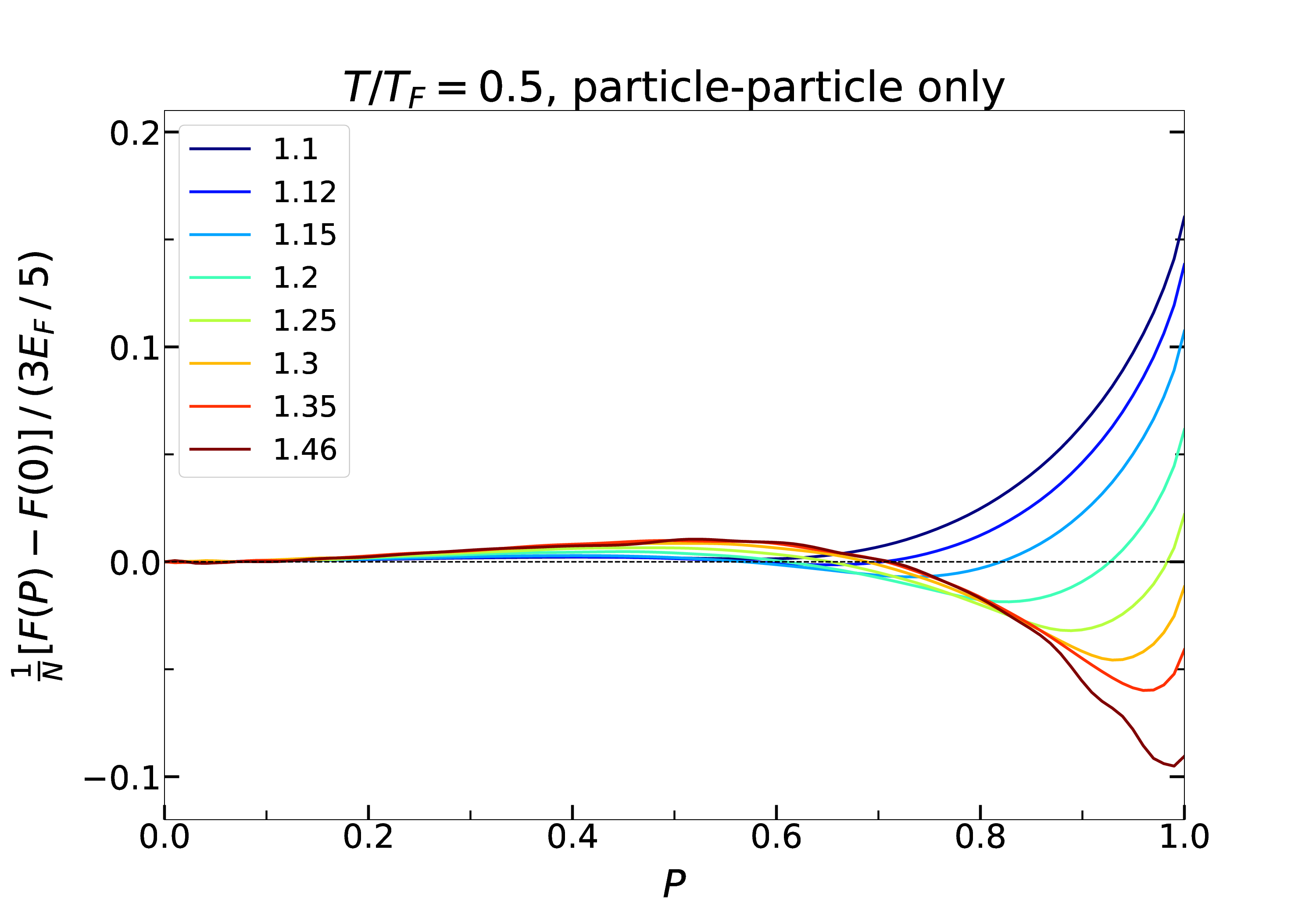,width=9.cm,height=7.0cm} 
}}
\caption{As in Figure \ref{fig:FresPPonlyT00and02} but for
  $T=0.3~\!T_{\rm F}$ and $T=0.5~\!T_{\rm F}$.}
\label{fig:FresPPonlyT03and05}
\end{figure}

\begin{figure}
\centerline{\hbox{
\psfig{figure=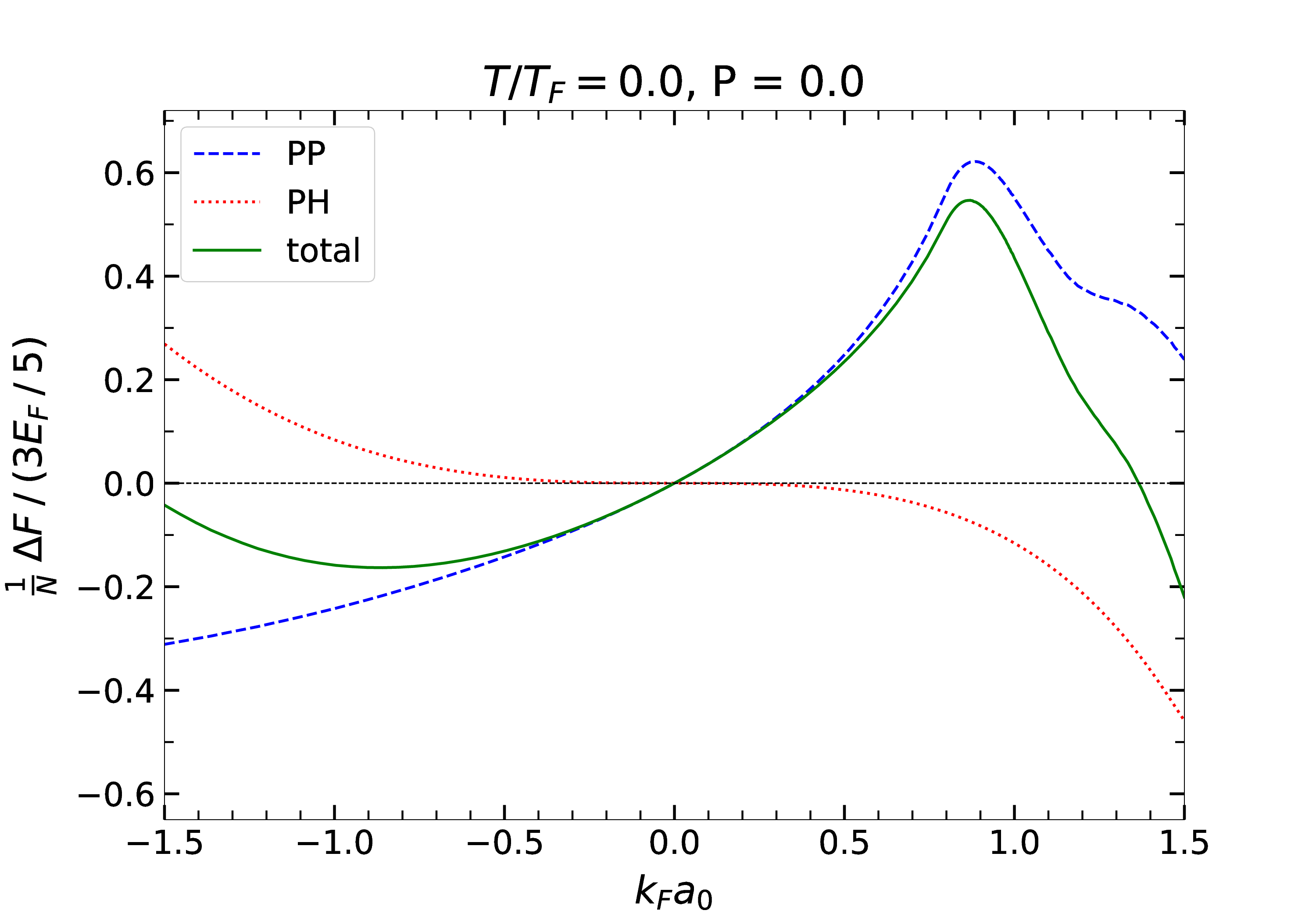,width=9.cm,height=7.0cm} 
\psfig{figure=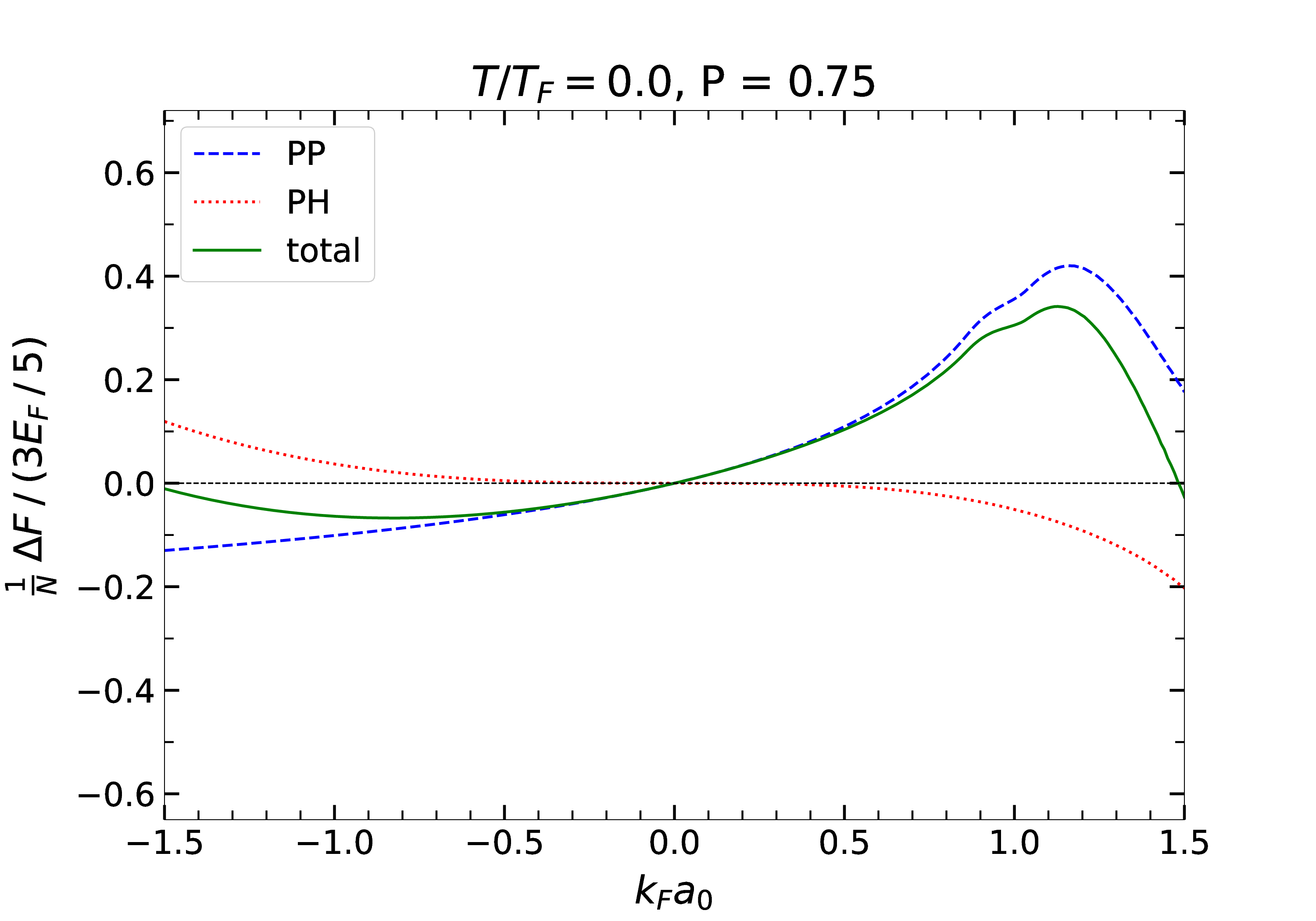,width=9.cm,height=7.0cm} 
}}
\caption{Dependence on the gas parameter $k_{\rm F}a_0$ of the resumed
  contributions of the particle-particle diagrams given by the expressions
  (\ref{eqn:FppSummed}) (dashed blue lines), of the particle-hole diagrams
  given by (\ref{eqn:FphSummed}) (dotted red lines) and of their sum (solid 
  green lines) for zero temperature and two values of the polarization $P$.} 
\label{fig:FakFT00}
\end{figure}

\begin{figure}
\centerline{\hbox{
\psfig{figure=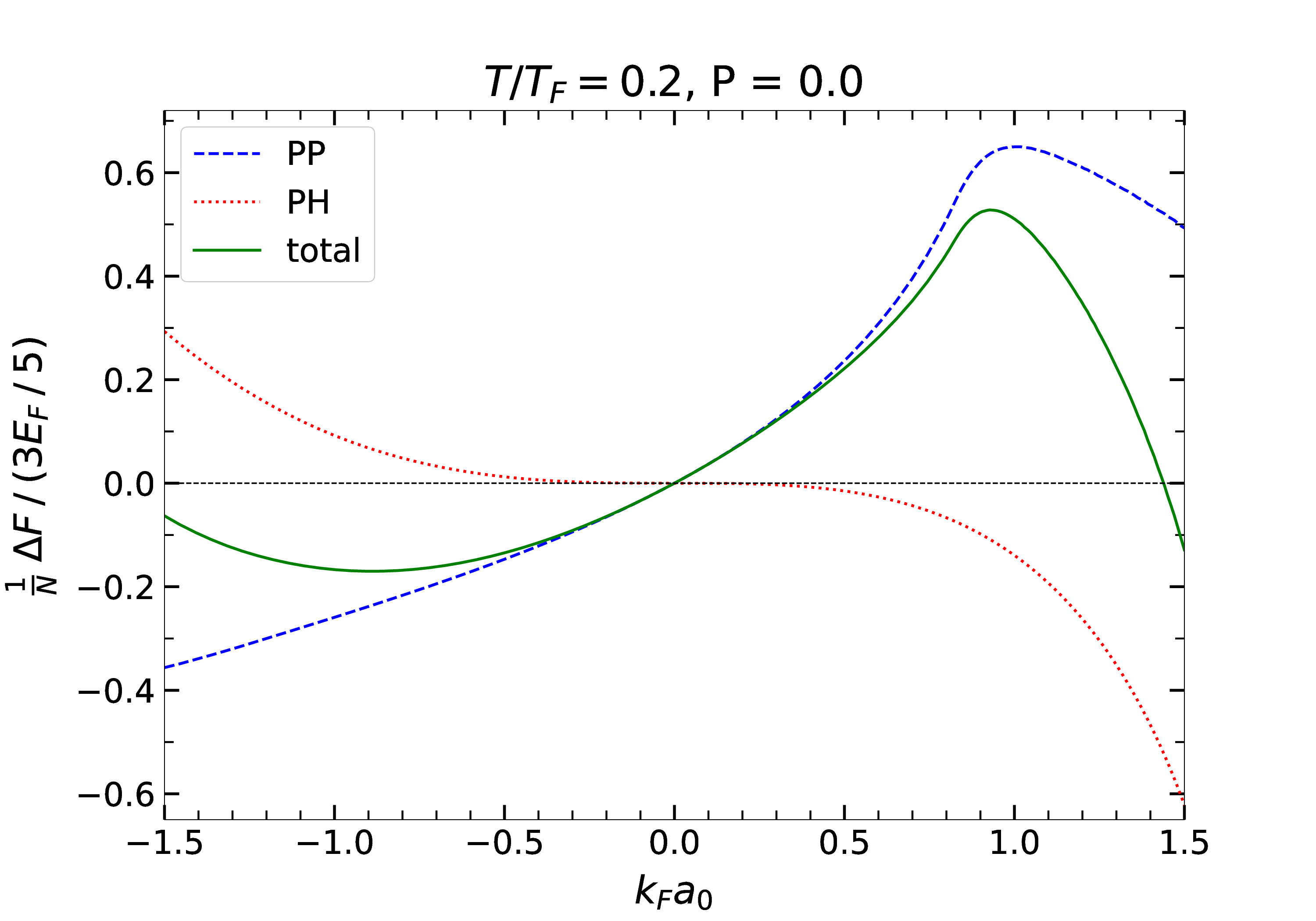,width=9.cm,height=7.0cm} 
\psfig{figure=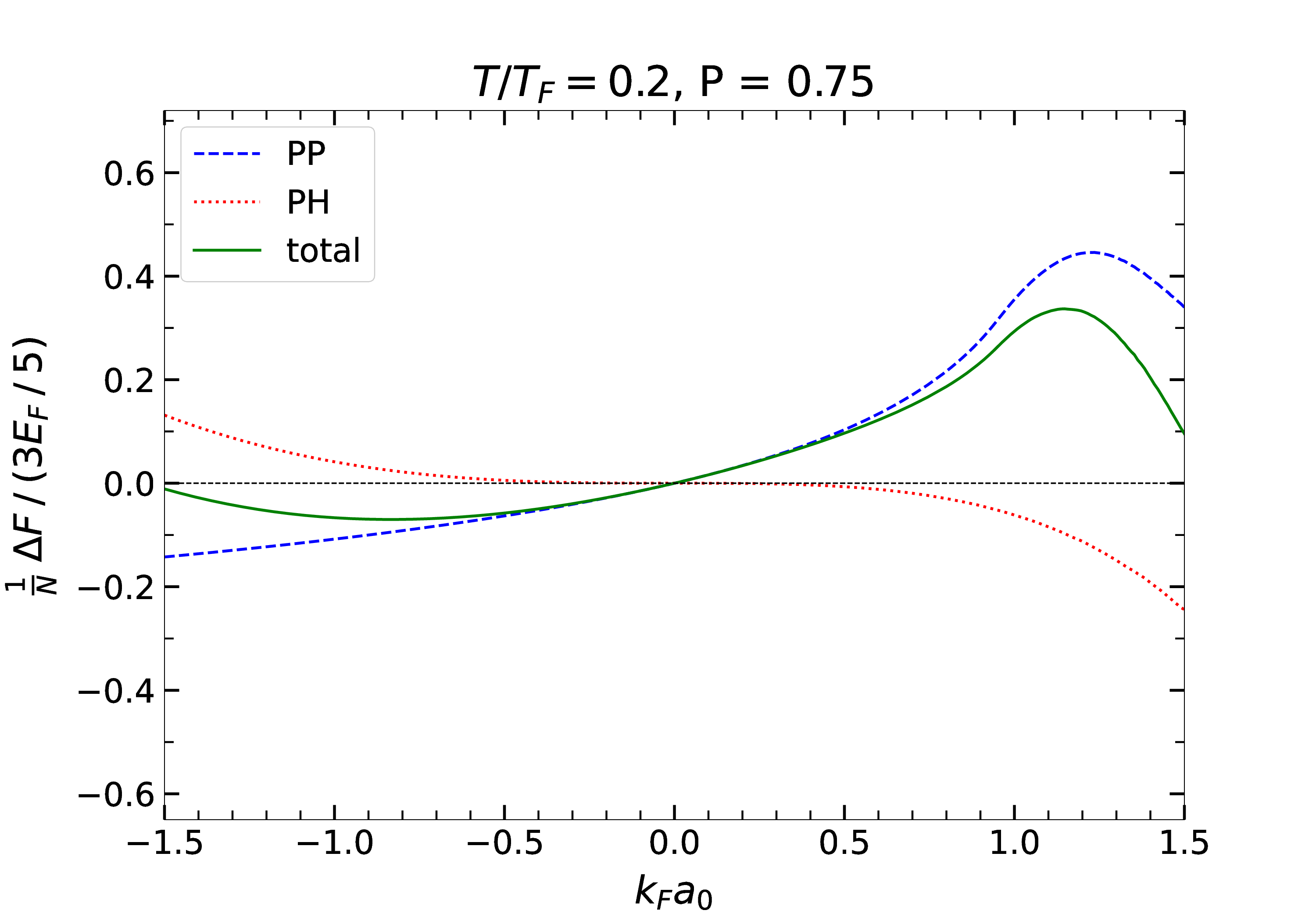,width=9.cm,height=7.0cm} 
}}
\caption{As in Figure \ref{fig:FakFT00} but for $T=0.2~\!T_{\rm F}$.}
\label{fig:FakFT02}
\end{figure}

\begin{figure}
\centerline{\hbox{
\psfig{figure=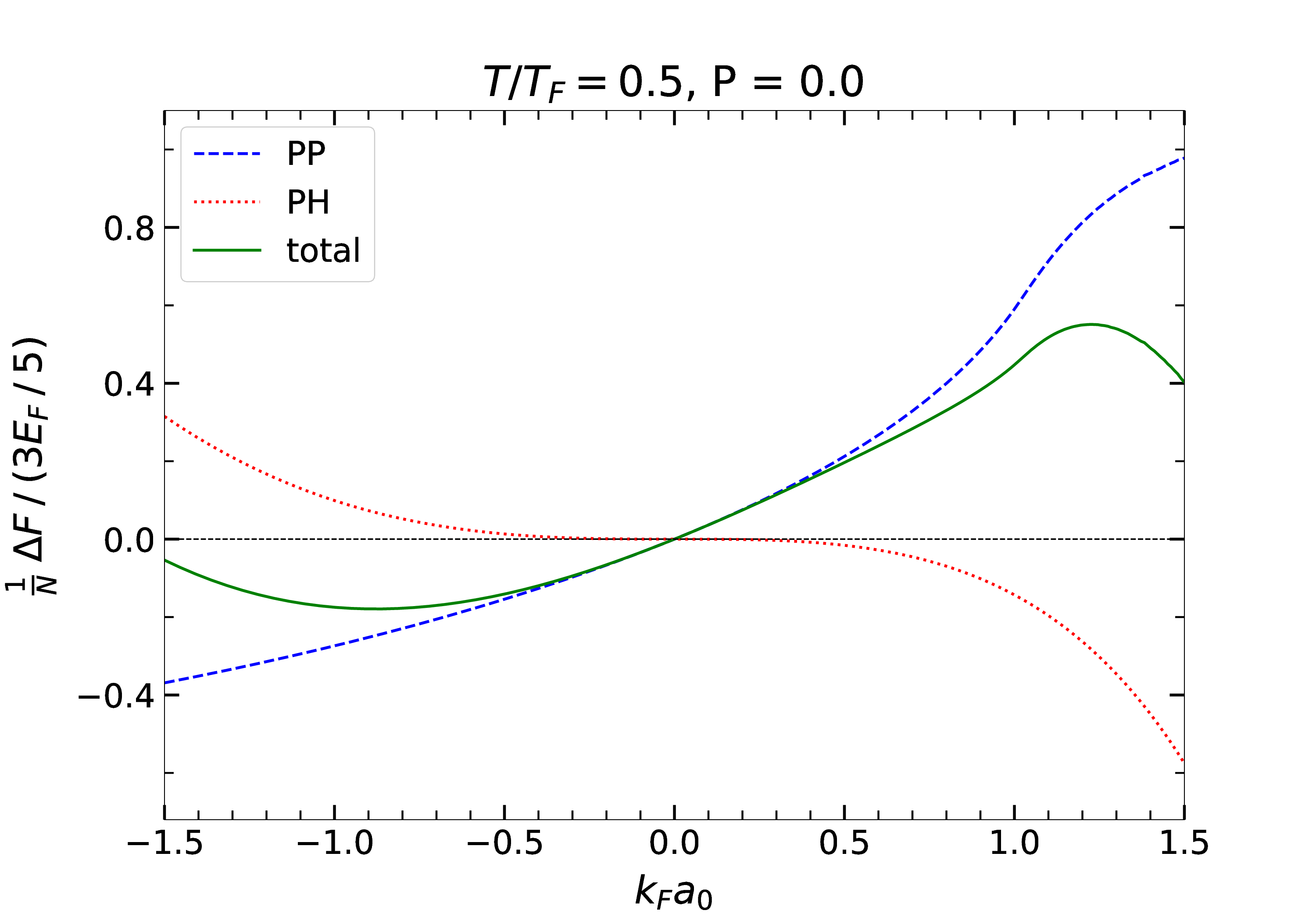,width=9.cm,height=7.0cm} 
\psfig{figure=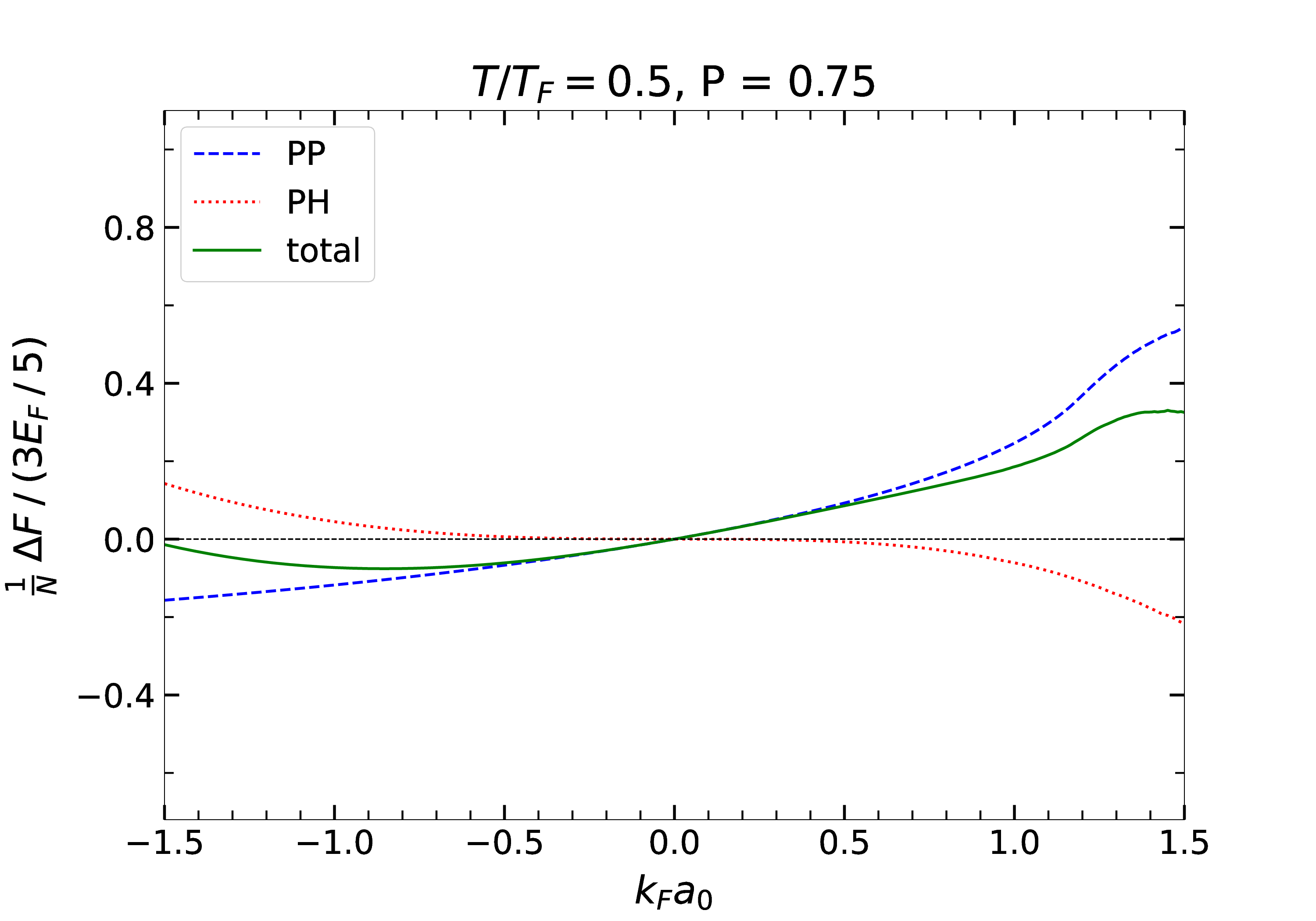,width=9.cm,height=7.0cm} 
}}
\caption{As in Figure \ref{fig:FakFT00} but for $T=0.5~\!T_{\rm F}$.}
\label{fig:FakFT05}
\end{figure}

The numerical procedure for evaluating the expression (\ref{eqn:F(2)final})
described in detail in Ref. \cite{CHGR} [the main trick is to construct -- for given
values of $t=k_{\rm B}T/\varepsilon_{\rm F}$ and $P$, which together, through
(\ref{eqn:nu(0)Determination}), determine the chemical potentials -- the
interpolations of the functions of the parameter $\Delta$ into which the
function (\ref{eqn:Lfunction}) can be decomposed] can be used to evaluate also
the expression (\ref{eqn:FppSummed}). Figures \ref{fig:FresPPonlyT00and02} and
\ref{fig:FresPPonlyT03and05} show for four different values of the temperature
and several values of the gas parameter $k_{\rm F}a_0$ the difference
$(F(P)-F(0))^{pp}/V$ [in units $(k^3_{\rm F}/3\pi^2)(3/5)\varepsilon_{\rm F}$]
obtained by adding to the zeroth order term (\ref{eqn:F(0)Term}) given
explicitly by the formula (25) in Ref. \cite{CHGR} the contribution
(\ref{eqn:FppSummed}). In agreement with the result obtained in the last
section of the work \cite{He1}, one can observe that for $T=0$ the minimum of
$(F(P)-F(0))^{pp}/V$ starts to move away from $P=0$ for $k_{\rm F}a_0=0.79$
indicating the continuous transition to the ordered state. As can be expected,
with increasing temperature this critical value of the expansion parameter
shifts toward larger values (0.85, 0.92 and 1.12 for $T=0.2~\!T_{\rm F}$,
$0.3~\!T_{\rm F}$ and $0.5~\!T_{\rm F}$, respectively) of the gas parameter
$k_{\rm F}a_0$. One can also see
that the minimum is back at $P=0$ for $k_{\rm F}a_0>0.96$ (at $T=0$) -- this is
the first order ``reentrant'' transition to the paramagnetic state observed in Refs.
\cite{He2,He3} which is the consequence of the existence of the maximum of
the energy density (at $T=0$) for $P=0$ treated as a function of $k_{\rm F}a_0$
shown by the dashed (blue) lines in the left panel of Figure \ref{fig:FakFT00}.
(From the physical point of view this reentrant transition is largely
irrelevant as it occurs for the values of the gas parameter for which the
formation of dimers prevails and the free energy of the metastable state is no
longer physical.) This maximum of the energy density (indeed seen in experiments
with cold gases \cite{MaxInEexp}), which occurs close to the Feshbach
resonance on its so-called BEC side (i.e. for large positive scattering
length $a_0$) and the existence of which for higher temperatures (for which
there is no phase transition) has been given a theoretical explanation (using a
completely different approach) in Ref. \cite{SheHo}, results here, as has been shown
in Refs. \cite{He1,He2}, from the appearance for $k_{\rm F}a_0>1.34$ of the simple
pole in the ``in-medium'' particle-particle elastic scattering amplitude
which can be interpreted as being due to the existence of the ``in-medium''
positive energy bound state of two fermions (of opposite spin projections).
The dashed (blue) lines on the right panel of this figure and in Figs.
\ref{fig:FakFT02} and \ref{fig:FakFT05} illustrate how the contribution
(\ref{eqn:FppSummed}) to the free energy and its maximum
change with the polarization and the temperature.

As the right panel of Fig. \ref{fig:FresPPonlyT00and02} and Fig.
\ref{fig:FresPPonlyT03and05} show, with increasing temperature the reentrant
transition occurs for higher ($1.06$ and $1.46$ for $T=0.2~\!T_{\rm F}$ and
$T=0.3~\!T_{\rm F}$, respectively and yet higher for $T=0.5~\!T_{\rm F}$) values
of the expansion parameter. The maximal depth of the minimum (at which $P\neq0$)
of $F$ first slightly decreases with the rising temperature (up to
$T\approx0.2~\!T_{\rm F}$) and then increases with it. Similarly $P=1$ is for
temperatures up to $T\approx0.2~\!T_{\rm F}$ reached only for $k_{\rm F}a_0$
values approaching the one at which the reentrant transition takes place but
for higher temperatures it is reached well before it.
\vskip0.5cm

\noindent{\large\bf 5. The particle-hole diagrams}
\vskip0.3cm

\noindent At the order $(k_{\rm F}a_0)^3$ to the free energy contributes also
the second diagram shown in Figure \ref{fig:C0cubeMercedes}. The corresponding
analytical expression is given by the convolution 
\begin{eqnarray}
  {F^{(3)ph}\over V}={C_0^3\over3}~\!{1\over\beta}\sum_l\!\int_{\mathbf q}
  [B(\omega^{\rm B}_l,{\mathbf q})]^3~\!,\label{eqn:F(3)phOriginal}
\end{eqnarray}
of three $B$-blocks which have the form \cite{CHGR}
\begin{eqnarray}
  B(\omega_l,{\mathbf q})=\int_{\mathbf p}{\{{\mathbf p}\}\over i\omega^{\rm B}_l
    -[{\mathbf p}]}~\!,\label{eqn:BblockExplicit}
\end{eqnarray}
analogous to (\ref{eqn:AblockExplicit}) but now with a different meaning
of the symbols $\{{\mathbf p}\}$ and $[{\mathbf p}]$:
\begin{eqnarray}
  &&N^{\mathbf p}_{+-}\equiv[1-n_+({\mathbf q}+{\mathbf p})]~\!n_-({\mathbf p})~\!,
  \nonumber\\
  &&N^{\mathbf p}_{-+}\equiv n_+({\mathbf q}+{\mathbf p})~\!
      [1-n_-({\mathbf p})]~\!,\label{eqn:BblockIds}\\
  &&\{{\mathbf p}\}\equiv n_+({\mathbf q}+{\mathbf p})-n_-({\mathbf p})
  =-N^{\mathbf p}_{+-}(1-e^{\beta[{\mathbf p}]})~\!,\nonumber\\
  &&[{\mathbf p}]\equiv\varepsilon_{\mathbf p}-\tilde\mu_-^{(0)}-
  \varepsilon_{{\mathbf q}+{\mathbf p}}+\tilde\mu_+^{(0)}~\!.\nonumber
\end{eqnarray}
After performing in (\ref{eqn:F(3)phOriginal}) the summation over the bosonic
Matsubara frequencies and using (\ref{eqn:BblockIds}) one arrives at
\begin{eqnarray}
  {F^{(3)ph}\over V}=-{C_0^3\over3}\!\int_{\mathbf q}\!\int_{{\mathbf p}_1}\!
  \int_{{\mathbf p}_2}\!\int_{{\mathbf p}_3}\!\left(N_{+-}^{{\mathbf p}_1}~\!
  {\{{\mathbf p}_2\}\over[{\mathbf p}_1]-[{\mathbf p}_2]}~\!
  {\{{\mathbf p}_3\}\over[{\mathbf p}_1]-[{\mathbf p}_3]}+{\rm two~other~terms}
  \right),\label{eqn:F(3)phSymmetric}
\end{eqnarray}
where ``two other terms'' means terms in which the role of $[{\mathbf p}_1]$
is played by $[{\mathbf p}_2]$ and $[{\mathbf p}_3]$. One can again make
contact with the order $(k_{\rm F}a_0)^3$ contribution of the particle-hole
diagram of Figure \ref{fig:C0cubeMercedes} to the ground state
energy density $E/V$ computed in \cite{CHWO3,CHWO4} where it was given
as a sum of two functions $K_1$ and $K_2$ (Eq. (17) in \cite{CHWO3}), by using
the algebraic identity
\begin{eqnarray}
N_{+-}^{{\mathbf p}_1}~\!
{N_{-+}^{{\mathbf p}_2}-N_{+-}^{{\mathbf p}_2}\over[{\mathbf p}_1]-[{\mathbf p}_2]}~\!
{N_{-+}^{{\mathbf p}_3}-N_{+-}^{{\mathbf p}_2}\over[{\mathbf p}_1]-[{\mathbf p}_3]}
+{\rm two~other~terms}\phantom{aaaaaaaaaaaaaaaaaaaaaaaaaaaaaaa}
\label{eqn:AlgIdB}\\
=N_{+-}^{{\mathbf p}_1}~\!
{N_{-+}^{{\mathbf p}_2}\over[{\mathbf p}_1]-[{\mathbf p}_2]}~\!
{N_{-+}^{{\mathbf p}_3}\over[{\mathbf p}_1]-[{\mathbf p}_3]}
+N_{-+}^{{\mathbf p}_1}~\!
{N_{+-}^{{\mathbf p}_2}\over[{\mathbf p}_1]-[{\mathbf p}_2]}~\!
{N_{+-}^{{\mathbf p}_3}\over[{\mathbf p}_1]-[{\mathbf p}_3]}
+{\rm two~other~terms},\nonumber
\end{eqnarray}
[it is in fact the  identity (\ref{eqn:AlgId}) but written with different
symbols]. Using the symmetry of this expression allows to write the expression
for $F^{(3)ph}/V$ as the sum of two terms that in the zero temperature limit
reproduce the two terms, $K_1$ and $K_2$, which in Refs. \cite{CHWO3,CHWO4}
represented the contribution of the second diagram of Fig.
\ref{fig:C0cubeMercedes} to $E/V$.

In order to properly treat the singularities in (\ref{eqn:F(3)phSymmetric})
we again make the substitutions
$[{\mathbf p}_l]\rightarrow[{\mathbf p}_l]+il\epsilon$ which allow to profit
from the symmetry of the three terms of (\ref{eqn:F(3)phSymmetric}). 
One then, similarly as  in the case of $F^{(3)pp}$, obtains
\begin{eqnarray}
  {F^{(3)ph}\over V}=-C_0^3\!\int_{\mathbf q}\!\int_{{\mathbf p}_1}\!
  N_{+-}^{{\mathbf p}_1}\!\left[\left({\rm P}\!\int_{{\mathbf p}_2}
  {\{{\mathbf p}_2\}\over[{\mathbf p}_1]-[{\mathbf p}_2]}\right)^2
  +{1\over3}\left(i\pi\!\!\int_{{\mathbf p}_2}\!\{{\mathbf p}_2\}~\!
  \delta([{\mathbf p}_1]-[{\mathbf p}_2])\right)^2\right].~
  \label{eqn:F3phCorrect}
\end{eqnarray}
Making now the changes of the variables: first
${\mathbf p}_1+{\mathbf q}={\mathbf k}_1$, and ${\mathbf p}_1=-{\mathbf k}_2$,
and then in the $n_+$ terms of $\{{\mathbf p}_2\}$ the change
${\mathbf p}_2+{\mathbf k}_1+{\mathbf k}_2=-{\mathbf p}$
and ${\mathbf p}_2={\mathbf p}$ in the $n_-$ terms of $\{{\mathbf p}_2\}$,
and then taking explicitly the integral over the cosine of the angle
between ${\mathbf p}$ and ${\mathbf k}_1+{\mathbf k}_2$
the expression for $F^{(3)ph}/V$ can be written in the form
\begin{eqnarray}
  {F^{(3)ph}\over V}=-C_0^{\rm R}\!\int_{{\mathbf k}_1}\!\int_{{\mathbf k}_2}\!
  [1-n_+({\mathbf k}_1)]~\!n_-({\mathbf k}_2)
  \left[K^2({\mathbf k}_1,{\mathbf k}_2)
    +{1\over3}\left(iK_\delta({\mathbf k}_1,{\mathbf k}_2)\right)^2\right],
  \label{eqn:F(3)phFinal}
\end{eqnarray}
in which the dimensionless functions $K$ and $K_\delta$ are given by the
integrals
\begin{eqnarray}
  K({\mathbf k}_1,{\mathbf k}_2)
  ={C_0^{\rm R}m_f\over(2\pi)^2\hbar^2|{\mathbf k}_1+{\mathbf k}_2|}
  \left(\int_0^\infty\!dp~\!p~\!n_+(p)
  \ln\!\left|{p-\Delta_1\over p+\Delta_1}\right|\right.\phantom{aaa}\nonumber\\
  \left.+\int_0^\infty\!dp~\!p~\!n_-(p)
  \ln\!\left|{p-\Delta_2\over p+\Delta_2}\right|\right),\label{eqn:Kfunction}
\end{eqnarray}
\begin{eqnarray}
  K_\delta({\mathbf k}_1,{\mathbf k}_2)
  =\pi~\!{C_0^{\rm R}m_f\over(2\pi)^2\hbar^2|{\mathbf k}_1+{\mathbf k}_2|}
  \left(\int_{|\Delta_1|}^\infty\!dp~\!p~\!n_+(p)
  -\int_{|\Delta_2|}^\infty\!dp~\!p~\!n_-(p)\right),\label{eqn:Kdeltafunction}
\end{eqnarray}
in which 
\begin{eqnarray}
  \Delta_1\equiv{{\mathbf k}_1\!\cdot\!({\mathbf k}_1+{\mathbf k}_2)\over
    |{\mathbf k}_1+{\mathbf k}_2|}~\!,\phantom{aaa}
 \Delta_2\equiv{{\mathbf k}_2\!\cdot\!({\mathbf k}_1+{\mathbf k}_2)\over
    |{\mathbf k}_1+{\mathbf k}_2|}~\!.
\end{eqnarray}
Again, the limits of the integrals in $K_\delta$ are determined by the condition
that the zeros of the arguments of the Dirac deltas treated as functions of
the cosine of the angle between ${\mathbf p}$ and
${\mathbf k}_1+{\mathbf k}_2$ lie between $-1$ and $+1$. And again
the function $K_\delta({\mathbf k}_1,{\mathbf k}_2)$ can be written
down in a closed analytical form - see below.

The expression (\ref{eqn:F(3)phFinal}) is finite although this is not
immediately obvious: while the integrals defining the functions $K$ and
$K_\delta$ are clearly (ultraviolet) finite owing to the presence of the
distribution functions $n_+(p)$ and $n_-(p)$, the integral over
$d^3{\mathbf k}_1$ has no such exponentially suppressing factor. In
addition, the presence of the factors $1/|{\mathbf k}_1+{\mathbf k}_2|$
in front of the functions $K$ and $K_\delta$ seem to imply the potential
presence of a
singularity in the integral over the directions of ${\mathbf k}_1$ (for
$|{\mathbf k}_1|=|{\mathbf k}_2|$). To prove the
finitness of (\ref{eqn:F(3)phFinal}) we will first analyze the behavior of the
difference $K^2-K_\delta^2/3$ in the limit $|{\mathbf k}_1|\rightarrow\infty$.
To this end it is helpful to split the functions $K$ and $K_\delta$ into
$K_++K_-$ and $K_{\delta+}+K_{\delta-}$ (the subscripts $\pm$ corresponds to the
$n_\pm$ distribution functions).
The hardest to see is the convergence of the integral over $d^3{\mathbf k}_1$
involving the factor $K_-^2-K^2_{\delta-}/3$. The dangerous term
$n_-({\mathbf k}_2)[K_-^2-K^2_{\delta-}/3]$ comes from the term
\begin{eqnarray}
\int_{\mathbf q}\!\int_{{\mathbf p}_1}\!\int_{{\mathbf p}_2}\!\int_{{\mathbf p}_3}
\left\{{n_-({\mathbf p}_1)~\!n_-({\mathbf p}_2)~\!n_-({\mathbf p}_3)\over
  ([{\mathbf p}_1]-[{\mathbf p}_2])([{\mathbf p}_1]-[{\mathbf p}_3])}
  +{\rm two~other~terms}\right\},\nonumber
\end{eqnarray}
in (\ref{eqn:F(3)phSymmetric}). This, however, is algebraically zero [just as
the integrand of (\ref{eqn:Integral})] and, therefore, the factor
$K_-^2-K^2_{\delta-}/3$ must be zero also. As any inaccuracy in the numerical
evaluation of the integrals (\ref{eqn:Kfunction}) and
(\ref{eqn:Kdeltafunction}) could lead to a nonzero difference 
$K_-^2-K^2_{\delta-}/3$ and thus to a (fake) nonconvergence of the
integration over $d^3{\mathbf k}_1$, in the term with unity arising from
$[1-n_+({\mathbf k}_1)]$ we simply replace $K^2-K^2_{\delta}/3$ by
$K_+^2+2K_+K_--(K_{\delta+}^2+2K_{\delta+}K_{\delta-})/3$. The rest of the terms
are separately integrable in the limit $|{\mathbf k}_1|\rightarrow\infty$:
\begin{eqnarray}
  K_{\delta+}\propto{1\over|{\mathbf k}_1+{\mathbf k}_2|}~\!
  \ln\!\left(1+e^{-\beta(\hbar^2\Delta_1^2/2m-\tilde\mu^{(0)}_+)}\right)
   \approx{1\over|{\mathbf k}_1+{\mathbf k}_2|}~\!
 e^{-\beta(\hbar^2\Delta_1^2/2m-\tilde\mu^{(0)}_+)}~\!,\nonumber
\end{eqnarray}
because $\Delta^2_1$ grows like ${\mathbf k}_1^2$ as
$|{\mathbf k}_1|\rightarrow\infty$. This secures the convergence of the
integrals of the terms $K_{\delta+}^2$ and $2K_{\delta+}K_{\delta-}$.
Similarly, it can be estimated that the integral over $p$ in
$K_+$ behaves as $1/\Delta_1$ when $|{\mathbf k}_1|\rightarrow\infty$.
Since each of the $K_\pm$ functions has the factor 
$1/|{\mathbf k}_1+{\mathbf k}_2|$ in front of it, the term $(K_+)^2$
behaves for $|{\mathbf k}_1|\rightarrow\infty$ as
$1/({\mathbf k}_1^2+{\mathbf k}_1\cdot{\mathbf k}_2)^2$ and this secures
the convergence of the integration over $d^3{\mathbf k}_1$. The term
$2K_+K_-$, instead, behaves only as
$1/({\mathbf k}_1^2+{\mathbf k}_1\!\cdot\!
{\mathbf k}_2)|{\mathbf k}_1+{\mathbf k}_2|$,
but the integration over the
cosine of the angle between ${\mathbf k}_1$ and ${\mathbf k}_2$ kills
the term of order $1/|{\mathbf k}_1|^3$ and the remaining terms are
already integrable. However, this only power-like suppression of the integration
of the term $K_+^2+2K_+K_-$ makes numerical evaluation of the
particle-hole contribution to the free energy more difficult and therefore
potentially less accurate than the evaluation of the particle-particle one.
As to the potentially singular factors $1/|{\mathbf k}_1+{\mathbf k}_2|$,
one should first notice that the original expressions (\ref{eqn:F(3)phOriginal})
and (\ref{eqn:BblockExplicit}) as well as similar formulas giving the
contributions of the $N$-th order particle-hole rings do not contain such
singularities. They are due to the symmetrizations needed to arrive at the final
formulas and must therefore cancel out [like the spurious singularities
of the integrand in (\ref{eqn:Integral})] even if it is not directly evident.
In the third order the finiteness of (\ref{eqn:F(3)phFinal}) can be seen as
follows: since the integrals in the definitions of the
$K_\pm$ and $K_{\delta\pm}$ functions are finite
in the limit ${\mathbf k}_1+{\mathbf k}_2\rightarrow{\mathbf0}$, the
singularities of (\ref{eqn:F(3)phFinal}) have essentially the form
$1/({\mathbf k}_1^2+{\mathbf k}_2^2+2|{\mathbf k}_1||{\mathbf k}_2|\xi)$ where
$\xi$ is the cosine of the angle between ${\mathbf k}_1$ and ${\mathbf k}_2$.
In the third order contribution considered here,
after integration over $\xi$ they give rise to terms
$\ln\!||{\mathbf k}_1|-|{\mathbf k}_2||$ which are
integrable.\footnote{In fact for $|{\mathbf k}_1|=|{\mathbf k}_2|$
  the factors $\Delta_1$ and $\Delta_2$ behave as $\sqrt{1+\xi}$ and
  it can be checked (numerically) that  the integrals in the functions $K_+$
  and $K_-$ vanish then as $\sqrt{1+\xi}$, so these functions have 
  finite limits. The singularities reside only in the terms
  involving the functions $K_{\delta\pm}$.}
In the numerical evaluation of the integrals in (\ref{eqn:F(3)phFinal})
and (\ref{eqn:FphSummed}) it is however sufficient to impose a cutoff
$|{\mathbf k}_1+{\mathbf k}_2|>\kappa$ and check that the results  stabilize
as  $\kappa$ approaches zero.
Thus the expression (\ref{eqn:F(3)phFinal}) is finite and we have checked that 
evaluated for $T=0$ (so that the Fermi distribution functions can be replaced
by the step functions) it reproduces numerically in the entire range of
polarizations $P$ the contribution to the ground state energy density
of the third order particle-hole diagram 
of Figure \ref{fig:C0cubeMercedes} obtained in \cite{CHWO3,CHWO4}.
\vskip0.1cm

Using the same tricks as previously the contribution to the free energy
of the infinite series of Feynman diagrams composed of $N$-fold rings of the
particle-hole $B$-blocks of Figure \ref{fig:ElementaryLoops} can be
summed in a closed form. The order $(k_{\rm F}a_0)^N$ term of this series is
\begin{eqnarray}
  {F^{(N)ph}\over V}=-(-1)^{N+1}~\!{C_0^N\over N}\!\int_{\mathbf q}\!
  \int_{{\mathbf p}_1}\!\dots\!\int_{{\mathbf p}_N}\sum_{n=1}^N
  N_{+-}^{{\mathbf p}_n}\left(\prod_{j\neq n}^N
  {\{ {\mathbf p}_j \}\over[{\mathbf p}_n]-[{\mathbf p}_j]
    +i(n-j)\epsilon}\right),
\end{eqnarray}
(apart from the extra minus originating from the minus sign in the identity
$\{{\mathbf p}\}=-N^{\mathbf p}_{+-}(1-e^{\beta[{\mathbf p}]})$, the origin of the
rest of the prefactor is the same as in the case of $F^{(N)pp}$) and, repeating
the steps one arrives at the formal sum
\begin{eqnarray}
  \sum_{N=1}^\infty{F^{(N)ph}\over V}
  =-~\!C_0^{\rm R}\int_{{\mathbf k}_1}\!\int_{{\mathbf k}_2}\!
  [1-n_+({\mathbf k}_1)]~\!n_-({\mathbf k}_2)
  ~\!{{\rm arctan}(K_\delta/(1-K))\over K_\delta}~\!.\label{eqn:FphSumFormal}
\end{eqnarray}
From this sum one has to subtract two first terms of the series:
there is no order $C_0^{\rm R}$ particle-hole diagram and the order
$(C_0^{\rm R})^2$ term  is already
(recall that the order $C_0^2$ contribution of the diagram shown in Fig.
\ref{fig:ElementaryLoops} can be written either as a convolution  of two
$A$-blocks or of two $B$-blocks shown in the same figure), in the properly
renormalized form, included in (\ref{eqn:FppSummed}). Thus the final
form of the resummed contributions of the particle-hole ring diagrams is
\begin{eqnarray}
{F^{ph}\over V}=-C_0^{\rm R}\int_{{\mathbf k}_1}\!\int_{{\mathbf k}_2}\!
[1-n_+({\mathbf k}_1)]~\!n_-({\mathbf k}_2)\left[
  {{\rm arctan}(K_\delta/(1-K))\over K_\delta}-1-K\right].\label{eqn:FphSummed}
\end{eqnarray}

One has to comment again on the finitness of the expression
(\ref{eqn:FphSummed}). The singularities related to the factors
$1/|{\mathbf k}_1+{\mathbf k}_2|$ are now harmless because the function $K$
is, as remarked, finite in the limit
${\mathbf k}_1+{\mathbf k}_2\rightarrow{\mathbf0}$ and the
singular function $K_\delta$ is now in the denominator and under the arctan
function. As to the ultraviolet finiteness
of (\ref{eqn:FphSummed}), the integral of the factor $K$ explicitly subtracted
in the square brackets is ultraviolet divergent being simply equivalent to the
divergent expression (\ref{eqn:F(2)inTermsOfC0}). To ensure proper cancellation
of this term in (\ref{eqn:FphSummed}) for large
$|{\mathbf k}_1|$ and/or $|{\mathbf k}_2|$ we expand the arctan function
in $K_\delta$ and $K$ up to the sixth order (we have checked that taking
more terms of the expansion does not change the result appreciably). The
ultraviolet divergent terms linear in $K$ then disappear and, moreover,
the expansion allows to implement the discussed trick with replacing
$K^2-K_\delta^2/3$ by $K^2_++2K_+K_--(K^2_{\delta+}+2K_{\delta+}K_{\delta-})$.
\vskip0.1cm

In Ref. \cite{CHWO4} we have compared the order $(k_{\rm F}a_0)^3$ contributions to
the ground state energy (i.e. to the free energy $F$ for zero temperature) of
the particle-particle and of the particle-hole diagrams (shown in Fig.
\ref{fig:C0cubeMercedes}) in the entire range of the polarization $P$ and
found that the second one is not much smaller than the first one.
In Figs. \ref{fig:FakFT00}-\ref{fig:FakFT05}
we compare the magnitudes of the resummed contributions (\ref{eqn:FppSummed})
and (\ref{eqn:FphSummed}) to the free energy as functions of $k_{\rm F}a_0$
for three different temperatures
and two representative values of the system's polarization $P$. We
also plot the sum of these two contributions. It again follows that for
$k_{\rm F}a_0\sim1$ the resummed contribution of the particle-hole diagrams
is not much smaller than that of the particle-particle ones
and is of the opposite sign. It is also clear that
in this most important region the sum of the two contributions is significantly 
distorted compared to the resummed contribution of the particle-particle
diagrams. This raises the question of what impact the resummed contribution
of the particle-hole ring diagrams has on the results reported in the
Refs. \cite{He1,He2,He3}.

\begin{figure}
\centerline{\hbox{
\psfig{figure=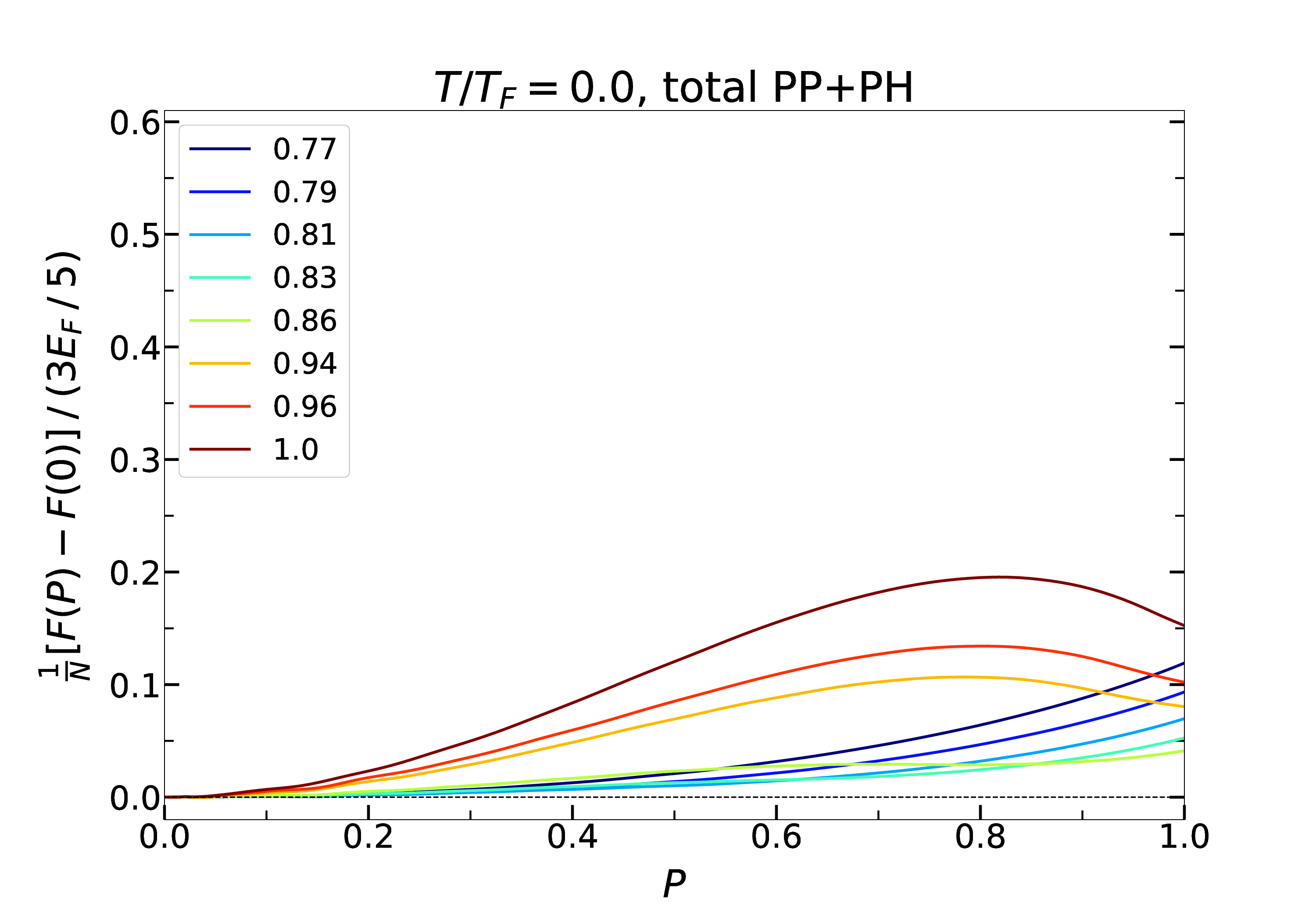,width=9.cm,height=7.0cm} 
\psfig{figure=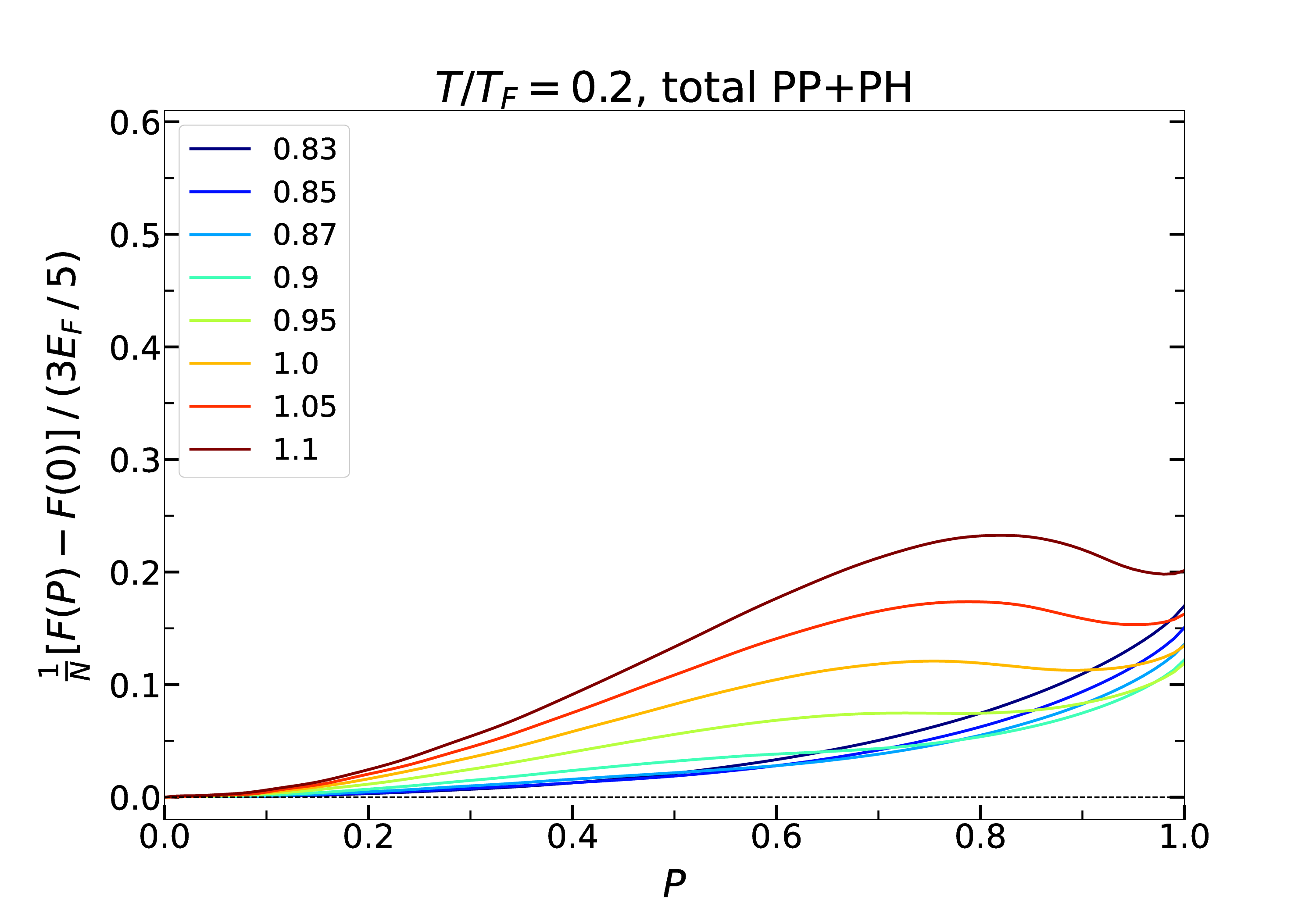,width=9.cm,height=7.0cm} 
}}
\caption{The difference $(F(P)-F(0))^{pp+ph}/N$ (in units
  $(3/5)\varepsilon_{\rm F}$ for $T=0$ and
  $T=0.2~\!T_{\rm F}$ as a function of the polarization
  $P=(N_+-N_-)/N$ for different values (indicated in the panels)
  of the gas parameter $k_{\rm F}a_0$.}
\label{fig:FresTotalT00and02}
\end{figure}

\begin{figure}
\centerline{\hbox{
\psfig{figure=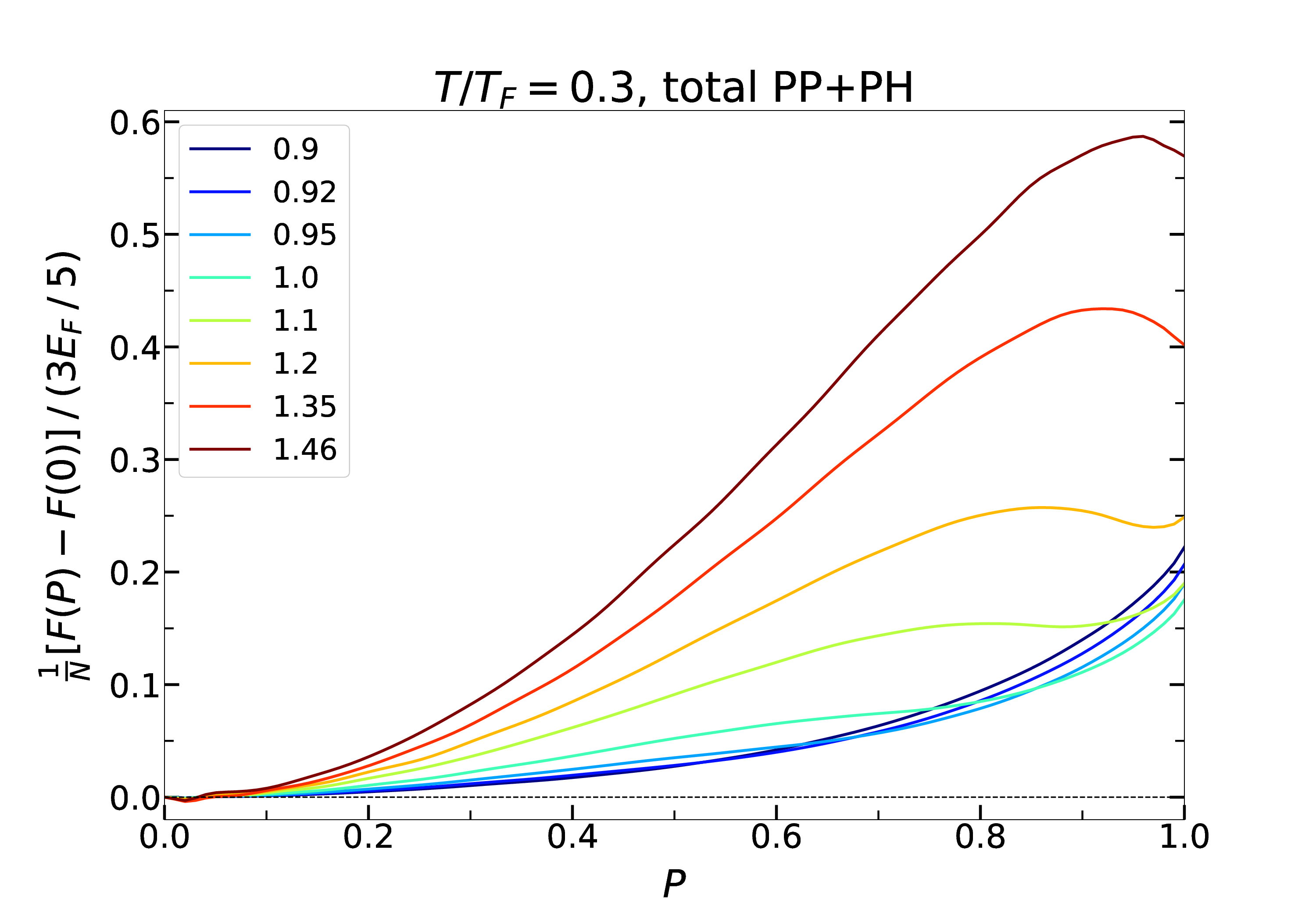,width=9.cm,height=7.0cm} 
\psfig{figure=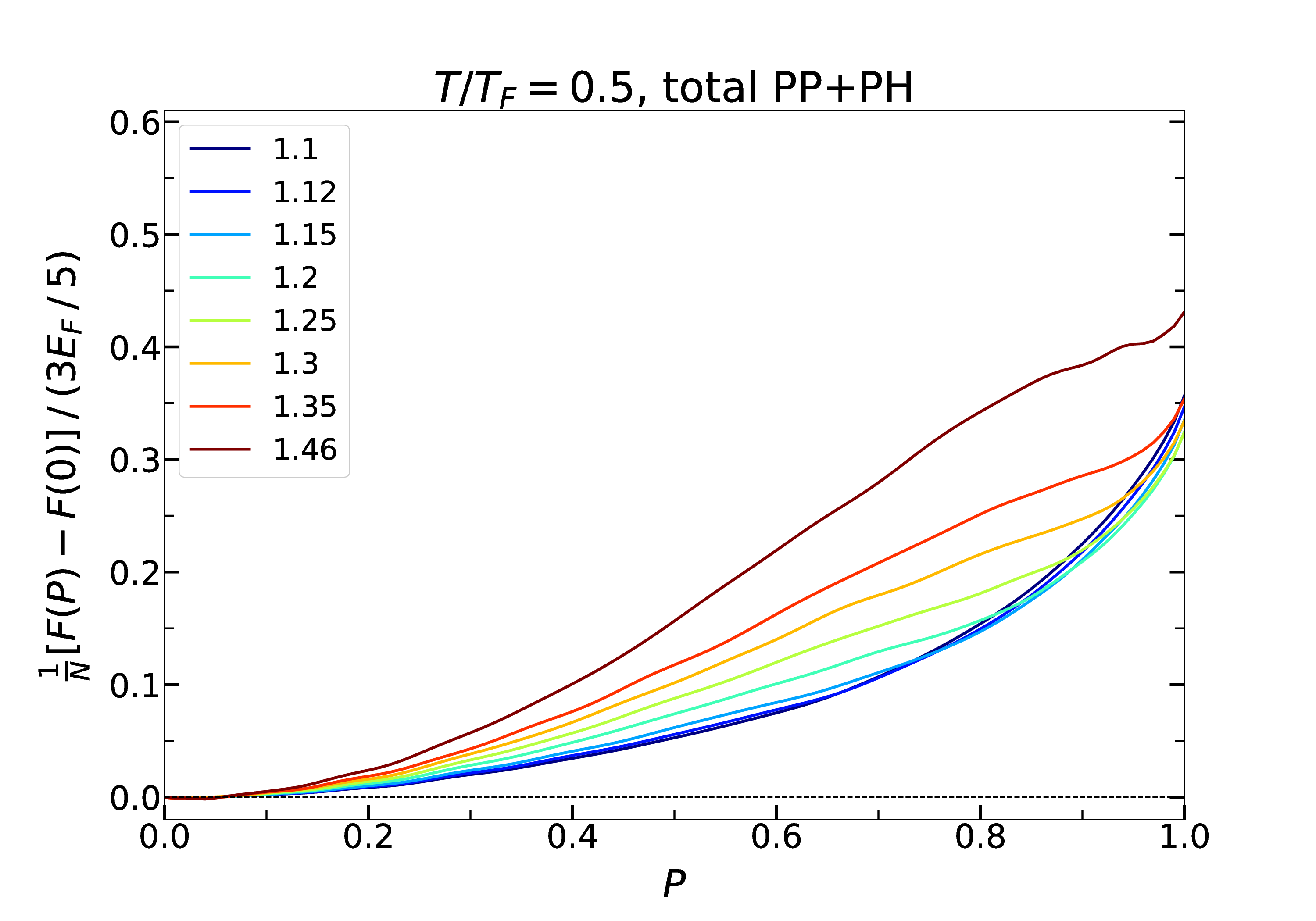,width=9.cm,height=7.0cm} 
}}
\caption{As in Figure \ref{fig:FresTotalT00and02} but for
  $T=0.3~\!T_{\rm F}$ and $T=0.5~\!T_{\rm F}$.}
\label{fig:FresTotalT03and05}
\end{figure}

Figures \ref{fig:FresTotalT00and02} and \ref{fig:FresTotalT03and05},
analogous to Figures \ref{fig:FresPPonlyT00and02} and
\ref{fig:FresPPonlyT03and05} illustrate the consequences of adding the
resummed contribution (\ref{eqn:FphSummed}) of the particle-hole 
ring diagrams to the free energy. The result is dramatic: the phase transition
to the ordered state discussed in Refs. \cite{He1,He2,He3} completely disappears!
Thus, even if the selection of the particle-particle ring diagrams as
giving the dominant contribution can be (partially) justified by invoking the
arguments, given in Ref. \cite{Steele}, based on using $1/2^{D/2}$ where $D$ is the
number of space dimensions, as the expansion parameter, they in practice do not
turn out to be really dominant: the contribution of other subsets of diagrams
(the number of such subsets beginning at a given order of the expansion
grows with the order number) can, as our results show, change qualitatively
the behavior of the thermodynamic potentials of the system of 
fermions close to the Feshbach resonance.

\vskip0.5cm

\noindent{\bf\large 7. Conclusions}
\vskip0.3cm

\noindent We have applied the systematic thermal (imaginary time) perturbative
expansion to the effective (low energy) field theory to compute the free
energy of the gas of interacting (nonrelativistic) spin $1/2$ fermions
for arbitrary values of the gas polarization and temperatures not exceeding the
Fermi temperature. We have shown how to circumvent the technical problem which
previously prevented us from immediately extending such a computation beyond
the second order in the gas parameter $k_{\rm F}a_0$ and have given explicit
formulas for the order $(k_{\rm F}a_0)^3$ contributions to the system's free
energy. It turned out that the analytical part of this computation 
is more transparent and easier than the corresponding
direct computation of the ground state energy based on the formula
(\ref{eqn:CorrectionsToE}) which gives only the zero temperature limit of the
results obtained with the help of the thermal expansion. (Of course, numerical
evaluation of the resulting expressions for a nonzero temperature is
considerably more involved than for $T=0$).

To obtain the complete order $(k_{\rm F}R)^3$ contribution to the free energy
of the gas of fermions interacting through a truly repulsive spin-independent
two-body potential (characterized by a length scale $R$) one would have to add
the contributions arising from the operators of lower length
dimension in the effective theory interaction term because
such potentials naturally give the $p$-wave scattering length $a_1$ and
the $s$-wave effective range $r_0$ comparable to the $s$-wave scattering
length $a_0$. Instead of doing this, in this work we have profited from the
simple structure of the contributions of the particle-particle and
particle-hole ring diagrams and managed to give simple formulas for their
contributions to the free energy resummed to all orders in the gas parameter
$k_{\rm F}a_0$. The obtained formulas therefore apply rather to cold gases of
fermionic atoms (interacting through attractive potentials)
close to the Feshbach resonance where their $s$-wave
scattering length $a_0$ is made positive and much larger than the remaining
scattering lengths and effective radii. Using these formulas we have
first checked
that including only the contributions of the particle-particle rings we
reproduce for zero temperatures all the results obtained in Ref. \cite{He1}.
In particular we confirm that in this approximation at zero temperature the
transition to the ordered phase occurs for $k_{\rm F}a_0=0.79$
and that it  is continuous. These results seem to agree
well with the results of the dedicated Monte Carlo computations.
Our formula would  allow to study also the thermal profile of the
transition, if it really existed. 

However we have found that the phase transition to the ordered state
completely disappears
after including into the free energy the resummed contribution of the
particle-hole ring diagrams - the minimum of the free energy is always for
zero polarization.
This may at least indicate that the agreement of the critical value of
the gas parameter $k_{\rm F}a_0$ found in Refs. \cite{He1,He2,He3}
with its value obtained from the Monte Carlo simulations (done with 
attractive potentials tuned so that $a_0$ is positive and large)
is just accidental. It may however also indicate that
there is indeed no transition (in the metastable state) to the magnetically ordered phase
if the true interaction is attractive (but $a_0$ is made positive and
large by approaching a Feshbach resonance).  The
vanishing at zero temperature of the inverse spin susceptibility
($1/\chi\propto(\partial^2F/\partial P^2)_{P=0}$) for some 
value of the gas parameter found in the Monte Carlo simulations
(which in the case of attractive potentials are not as clean as in the
repulsive case because of the need to exclude - in the finite volume and a
finite number of fermions, that
is without taking the thermodynamic limit  - an overlap with the true
ground state of the considered system of atoms) \cite{QMC10} from which one
infers its existence and its continuous character would then be
misleading (the second order approximation to $F$ predicts
analytically vanishing of the
inverse spin susceptibility at $k_{\rm F}a_0=1.058$ whereas the transition
is first order and occurs for $k_{\rm F}a_0=1.054$ \cite{He1}).
If this hypothesis is true it would provide yet
another (in addition to the formation of atomic dimers) reason for the
failure to simulate the itinerant ferromagnetism with the help of cold atoms.
In an case it is clear that more theoretical studies are needed to clarify
the situation. An obvious step in this direction would be extending
the computation of ref. \cite {WeDrSch} of the order $(k_{\rm F}a_0)^4$
contribution to nonequal densities of spin up and spin down fermions.
\\
\begin{center}
\textbf{Data Availability Statement}
\\
The source code for this project is publicly available on Zenodo  \cite{zenodo}.
\end{center}

\vskip0.5cm

\begin{center}
\textbf{Acknowledgements}
\\
This research was funded, in whole or in part, by l’Agence Nationale de la Recherche (ANR), project ANR-23-CE31-0019..
\end{center}


\vskip0.5cm

\begin{thebibliography}{99}
\bibitem{Lenz} W. Lenz, {\em  Z. Phys.} {\bf 56}, 778 (1929).
\bibitem{Stoner} E. Stoner, {\em The London, Edinburgh and Dublin
  Phil. Mag. and Journal of Science} {\bf15} (1933), 1018.
\bibitem{HuangYang57} K. Huang and C.N. Yang, {\em Phys. Rev.} {\bf 105},
   767 (1957), T.D. Lee and C.N. Yang, {\em Phys. Rev.} {\bf 105}, 1119 (1957).
\bibitem{Kesio} K. Huang, {\it Statistical Mechanics}, John Willey and Sons,
  Inc., New York 1963.
\bibitem{Pathria} R.K. Pathria, {\it Statistical Mechanics}, Pergamon Press,
  Oxford, 1972.
\bibitem{FetWal} A.L. Fetter and J.D. Walecka,
  {\it Quantum Theory of Many Particle Systems}, McGraw Hill, 1971.
\bibitem{KolczastyiSka} see e.g. Proceedings of the Joint Caltech/INT Workshop
  {\it Nuclear Physics with Effective Field Theory}, ed. R. Seki, U. van Kolck
  and M. Savage (World Scientific, 1998); Proceedings of the INT Workshop
  {\it Nuclear Physics with Effective Field Theory II}, ed. P.F. Bedaque,
  M. Savage, R. Seki and U. van Kolck (World Scientific, 2000);
  R.J. Furnstahl and H.-W. Hammer, {\em Phys. Lett.}
  {\bf B531}, 203 (2002); H.-W. Hammer, S. K\"onig and U. van Kolck,
  {\em Rev. of Mod. Phys.} {\bf 92} (2020), 025004;
\bibitem{HamFur00} H.-W. Hammer, R. J. Furnstahl, {\em Nucl. Phys.} {\bf A678},
  277 (2000); {\sf arXiv:0004043 [nucl-th]}.
\bibitem{HamFur02} R.J. Furnstahl, V.
  Steele and N. Tirfessa, {\em Nucl. Phys.} {\bf A671} (2000), 396;
  R.J. Furnstahl, H.-W. Hammer and N. Tirfessa,
  {\em Nucl. Phys.} {\bf A689},  846 (2001).
\bibitem{WeDrSch} C. Wellenhofer, C. Drischler and A. Schwenk,
  {\em Phys. Lett.} {\bf B802} 135247 (2020), {\sf arXiv:1812.08444 [nucl-th]};
  {\em Phys. Rev.} {\bf C104}, 014003 (2021),
  {\sf arXiv:2102.05966 [cond-mat.quant-gas]}.
\bibitem{CHWO1} P.H. Chankowski and J. Wojtkiewicz, {\em Phys. Rev.}
  {\bf B104} 144425 (2021), {\sf arXiv:2108.00793 [cond-mat.quant-gas]}. 
\bibitem{KANNO} S. Kanno, {\em Prog. Theor. Phys.} {\bf 44}, 813 (1970).
\bibitem{PECABO1} J. Pera, J. Casulleras and J. Boronat, {\em SciPost Phys.}
  {\bf 14}, 038 (2023), {\sf arXiv:2205.13837 [cond-mat.quant-gas]}.
\bibitem{CHWO3} P.H. Chankowski, J. Wojtkiewicz and R. Bakhshizada,
  {\em Acta. Phys. Pol.}
  {\bf B53} (2022), 9-A4, {\sf arXiv:2206.05076 [cond-mat.quant-gas]}.
\bibitem{CHWO4} P.H. Chankowski, J. Wojtkiewicz and S. Augustynowicz,
  {\em Phys. Rev.}
  {\bf A107} (2023) 063311, {\sf arXiv:2303.09921 [cond-mat.quant-gas]}. 
\bibitem{PECABO2} J. Pera, J. Casulleras and J. Boronat, {\em SciPost Phys.}
  {\bf 17}, 030 (2024), {\sf arXiv:2206.06932 [cond-mat.quant-gas]};
   {\sf arXiv:2407.14137 [cond-mat.quant-gas]}.
\bibitem{BeKiVoj} D. Belitz, T.R. Kirkpatrick and T. Vojta,
  {\em Phys. Rev. Lett.} {\bf 82}, 4707 (1999).
\bibitem{He1} L. He and X.-G. Huang, {\em Phys. Rev.} {\bf A85}, 043624 (2012),
  {\sf arXiv:1106.1345}.
\bibitem{He2} 
  L. He, {\em Ann. of Phys.} {\bf 351}, 477 (2014), {\sf arXiv:1405.3338}.
\bibitem{HEIS} H Heiselberg, {\em Phys. Rev.} {\bf A83} (2011), 053635.
\bibitem{QMC10} S. Pilati, G. Bertaina, S. Giorgini and M. Troyer,
  {\em Phys. Rev. Lett.} {\bf 105}, 030405 (2010),
  {\sf arXiv:1004/1169 [cond-mat.quant-gas]}.
\bibitem{ChiGriJuTie} C. Chin, R. Grimm, P. Julienne and E. Tiesinga,
  {\em Rev. Mod. Phys.} {\bf 82}, 1225 (2010).
\bibitem{Pippard} A.B. Pippard, {\it The Elements of Classical Thermodynamics},
  Cambridge University Press, 1964.
\bibitem{ItFMObs} G.-B. Jo et al.,
{\em Science} {\bf 325}, 1521 (2009).
\bibitem{ItFMNotObsT} D. Pekker et al., {\em Phys. Rev. Lett.} {\bf106}
  (2011), 050402.
\bibitem{ItFMNotObsE}
  Y.-R. Lee et al., {\em Phys. Rev.} {\bf A85} (2012), 063615;
  C. Sanner et al., {\em Phys. Rev. Lett.} {\bf108} (2012), 240404.
\bibitem{DUMacDO} R.A. Duine and A.H. MacDonald, {\em Phys. Rev. Lett.}
  {\bf 95}, 230403 (2005).
\bibitem{LL} Landau L.D. and Lifshitz E.M., {\it Statistical Physics}, 3rd ed.,
  Pergamon Oxford 1980.
\bibitem{CHGR} O. Grocholski and P.H. Chankowski, {\em Acta Phys. Pol.}
  {\bf B54} (2023), 11-A4, {\sf arXiv:2308.14782 [cond-mat.quant-gas]}.
\bibitem{He3} L. He, {\em Phys. Rev.} {\bf A90}, 053633, {\sf arXiv:1405.5242}.
\bibitem{KAJZERKA1} N. Kaiser, {\em Nucl. Phys.} {\bf A860}, 41 (2011),
         {\sf arXiv:1102.2154}.
\bibitem{KAJZERKA2} N. Kaiser, {\em Eur. Phys. J.} {\bf A48}, 148 (2012),
  {\sf arXiv:1210.0783}.
\bibitem{MaxInEexp} T. Bourdel et al. {\em Phys. Rev. Lett.} {\bf 91},
  020402 (2003); B. Fr\"olich  et al. {\em Phys. Rev. Lett.} {\bf 106},
  105301 (2011).
\bibitem{SheHo} V.B. Shenoy and T.-L. Ho, {\em Phys. Rev. Lett.} {\bf 107},
  210401 (2011).
\bibitem{Steele}  J.V. Steele, {\sf arXiv:0010066 [nucl-th]}; T. Sch\"afer,
  C.-W. Kao, and S.R. Cotanch,  {\em Nicl. Phys.} {\bf A762}, 82 (2005).
\bibitem{zenodo} O. Grocholski, \url{https://doi.org/10.5281/zenodo.18892299}.
  
\end{thebibliography}
\end{document}